\documentclass[prd,10pt,aps,twocolumn,nofootinbib,floatfix]{revtex4-1}
\usepackage{combelow}
\usepackage{mathtools}
\usepackage{footmisc}
\usepackage{amsmath}
\usepackage{amssymb}
\usepackage{slashed}
\usepackage{epsfig}
\usepackage{xcolor}
\usepackage{pifont}

\newcommand{\beqn}{\begin{eqnarray}}
\newcommand{\eeqn}{\end{eqnarray}}
\newcommand{\littlecross}{\scriptsize{\mbox{\ding{59}}}}
\newcommand{\beqs}{\begin{subequations}}
\newcommand{\eeqs}{\end{subequations}\\[-2mm]\noindent}
\newcommand{\eq}[1]{(\ref{#1})}

\newcommand{\cD}{{\cal D}}
\newcommand{\sumint}[1]{\sum \hskip -5mm\int\limits_{#1} \hskip 2mm}
\newcommand{\sumintt}[1]{\Sigma_{#1}\hskip -5mm\int \hskip 2mm}

\newcommand{\Z}{{\mathbb Z}}
\newcommand{\R}{{\mathbb R}}
\newcommand{\Dirac}{\slashed \cD}

\newcommand{\bs}{\boldsymbol}

\newcommand{\ket}[1]{{\left|#1\right\rangle}}
\newcommand{\avr}[1]{{\left\langle #1 \right\rangle}}

\definecolor{purple}{rgb}{0.8,0,0.6}

\allowdisplaybreaks
\begin{document}

\title{Fractal thermodynamics and ninionic statistics of coherent rotational states: realization via imaginary angular rotation in imaginary time formalism}

\author{M. N. Chernodub}
\affiliation{Institut Denis Poisson UMR 7013, Universit\'e de Tours, 37200, Tours, France}

\begin{abstract}
We suggest the existence of systems in which the statistics of a particle changes with the quantum level it occupies. The occupation numbers in thermal equilibrium depend on a continuous statistical parameter that interpolates between bosonic or fermionic and ghost-like statistical distributions. We call such particle states ``ninions'': they are different from anyons and can exist in 3+1 dimensions. We suggest that ninions can be associated with coherent angular momentum states. In the Euclidean imaginary-time formalism, the ninionic statistics can be implemented via the rotwisted boundary conditions, which are associated with the rigid global rotation of the system with an imaginary angular frequency. The imaginary rotation is characterized by a $PT$-symmetric non-Hermitian Hamiltonian and possesses a well-defined thermodynamic limit. The physics of ninions in thermal equilibrium is accessible for numerical simulations on Euclidean lattices. We provide a no-go theorem on the absence of analytical continuation between real and imaginary rotations in the thermodynamic limit. The ground state of ninions shares similarity with the $\theta$-vacuum in QCD. The ninions can produce negative pressure and energy, similar to the Casimir effect and the cosmological dark energy. In the thermodynamic limit, the dependence of thermal energy of free ninions on the statistical parameter is a fractal.
\end{abstract}

\date{\today}

\maketitle

\section{Introduction}

The spin-statistics theorem implies that, in three spatial dimensions, local particle fields can possess only two types of statistics. Integer spins are associated with bosonic particles characterized by commuting fields, while half-integer spins describe fermionic particles described by anti-commuting fields~\cite{Finkelstein:1968hy}. In two spatial dimensions, the spin-statistics theorem does not work, and the value of particle spin can take any value~\cite{Leinaas:1977fm} thus supporting the existence of the third kind of particles, the anyons~\cite{PhysRevLett.49.957}. The statistics of particles is of crucial importance as it plays a principal role in the physical properties of any many-particle systems.

The rotation is at the heart of the statistical properties of particles. Under the full $2\pi$ rotation, the boson wavefunction stays the same; the fermion wavefunction picks a minus sign while the anyon wavefunction gets multiplied by the phase $e^{i \theta}$ which interpolates between the bosonic, $\theta = 0$ and fermionic, $\theta = \pi$ cases (and therefore one often says that the anyons possess the fractional statistics). The same phases appear under the exchange of two indistinguishable particles. 

This paper considers imaginary rotation, which corresponds to quantum systems that rotate with imaginary angular frequency. The imaginary rotation can naturally be formulated in the Euclidean imaginary-time formalism, which describes thermodynamic systems residing in thermal equilibrium\footnote{The Euclidean formulation of a theory is usually obtained after a Wick transformation which is also called ``the Wick rotation''. To avoid confusion with the main topic of this article, we use the term ``the Wick transformation''.}. The imaginary rotation serves as an analytical tool often used in description of thermodynamics of real rotating quantum systems~\cite{Yamamoto:2013zwa,Braguta:2020biu,Braguta:2021jgn,Chernodub:2020qah,Chen:2022smf,Chernodub:2022wsw,Chernodub:2022veq}.

Here we concentrate on the effects of the rotation with imaginary frequency on particle statistics. In Section~\ref{sec_imaginary_rotation} we describe how the imaginary rotation can be formulated in terms of a simple rotwisted boundary condition in the compactified imaginary time direction (Section~\ref{sec_imaginary_boundary}). We also show that the imaginary rotation introduces the statistical parameter $\chi$, which stipulates the surprising fractal properties of thermodynamics (Section~\ref{sec_fractal_thermodynamics}). The latter fact leads to a no-go theorem for analytical continuation from imaginary to real angular frequencies (Section~\ref{sec_no_go}) if the rotation is understood as a boundary condition. Section~\ref{sec_transmutation} demonstrates that the imaginary rotation leads to the statistical transmutations of particles and the appearance of a new, ninionic-type of statistics, which is different from anyons. In Section~\ref{sec_coherent}, we argue that the ninions are the coherent states of the angular momentum operator, which share similarity with the coherent spin states and coherent angular momentum stats~\cite{atkins1971angular} that, in turn, are similar to the coherent states of the linear harmonic oscillator~\cite{radcliffe1971some}. The coherent spin states play an important role in quantum optics~\cite{arecchi1972atomic}. The last Section is devoted to our conclusions.

\section{Imaginary rotation: thermodynamics}
\label{sec_imaginary_rotation}

\subsection{Imaginary rotation as a boundary condition}
\label{sec_imaginary_boundary}

Consider a quantum-mechanical system of bosonic or fermionic particles which rotates rigidly with the constant angular velocity ${\bs \Omega} = \Omega {\bs n}$ about the axis $\bs n$. Due to the rigid nature of rotation, the maximal spatial size $R$ of the system must be bounded in the $xy$ plane, $|\Omega| R < 1$, in order to avoid a clash with causality~\cite{Davies:1996ks,Ambrus:2014uqa}. The system resides in thermal equilibrium characterized by temperature $T = 1/\beta$, defined, along with the chemical potential~$\mu$, in the frame, which co-rotates with the system. In the co-rotating frame, the thermal parts of bosonic and fermionic free energies take, respectively, the following forms:
\beqn
&& \hskip -9mm 
F^{\mathrm{(b)}}_\beta {=} \frac{V}{2\beta} \sumint{\alpha,m} \!\!\!\! 
\sum_{c,r= \pm 1}\! \ln \left( 1 - e^{- \beta (\omega_{\alpha,m} - r\mu - c m \Omega)} \right)\!,
\label{eq_F_bos_rot} \\
&& \hskip -9mm
F^{\mathrm{(f)}}_\beta {=} \frac{V}{2\beta} \sumint{\alpha,m} \!\!\!\! 
\sum_{c,r= \pm 1} \!\!
\ln \left( 1 + e^{- \beta (\omega_{\alpha,m} - r\mu - c (m + \frac{1}{2}) \Omega)} \right)^{\!-1}\!\!\!,
\label{eq_F_ferm_rot}
\eeqn
where $V$ is the spatial volume of the system, $\omega = \omega_{\alpha,m}$ is the energy spectrum of the particles, and $\alpha$ is a collective notation of quantum numbers other than the projection of angular momentum, $m \in \Z$, on the axis of rotation~$\bs n$. Without loss of generality, we will assume that $\bs n$ is directed along the $z$ axis while using from time to time $\bs n$ to keep the generality of the expressions. For Dirac fermions, the index $\alpha$ includes also spin polarizations, $s_z = \pm 1/2$. The particle/anti-particle branches are represented by the index $r$. Throughout the paper, we work with units $\hbar = c = k_B = 1$.

The free energies also contain the zero-point contributions, $F^{\mathrm{(b)}}_0 = V f_0$ for bosons and  $F^{\mathrm{(f)}}_0 = - 2 V f_0$ for fermions, with $f_0 = \frac{1}{2} \sumintt{\alpha} \omega_{\alpha,m}$. These quantities depend neither on chemical potential, temperature, or angular frequency $\Omega$; therefore, we will ignore them below.

In the standard imaginary time formalism, the Wick transformation substitutes the time variable $t$ by the imaginary time $\tau = i t$. Thus, the quantum theory in thermal equilibrium is formulated in Euclidean space with coordinates $({\bs x},\tau)$. At finite temperature $T$, the imaginary time direction is compactified to a circle of the length $\beta = 1/T$. 

The partition functions~\eq{eq_F_bos_rot} and \eq{eq_F_ferm_rot} can be represented as traces over all quantum states of the statistical density matrix, ${\mathcal Z} = {\mathrm{Tr}} \, e^{- \beta (\hat{\mathcal H}_{{\bs \Omega}} - \mu \hat{\mathcal N})}$, which includes the Hamiltonian in the co-rotating frame:
\beqn
\hat{\mathcal H}_{{\bs \Omega}} = \hat{\mathcal H}_0 - \hat {\bs \Omega} \cdot \hat{\bs {\mathcal J}}\,,
\label{eq_H_Omega}
\eeqn
where $\hat{\mathcal H}_0$ is the Hamiltonian in the laboratory frame. The spectrum of the Hamiltonian~\eq{eq_H_Omega}, both for fermionic and bosons systems, is bounded from below provided the causality constraint is respected, $|\Omega| R < 1$. The corresponding partition function, at real angular frequency, is a well-defined quantity in the Euclidean imaginary time formalism.

\begin{figure}[!thb]
\centerline{\includegraphics[scale=0.5,clip=true]{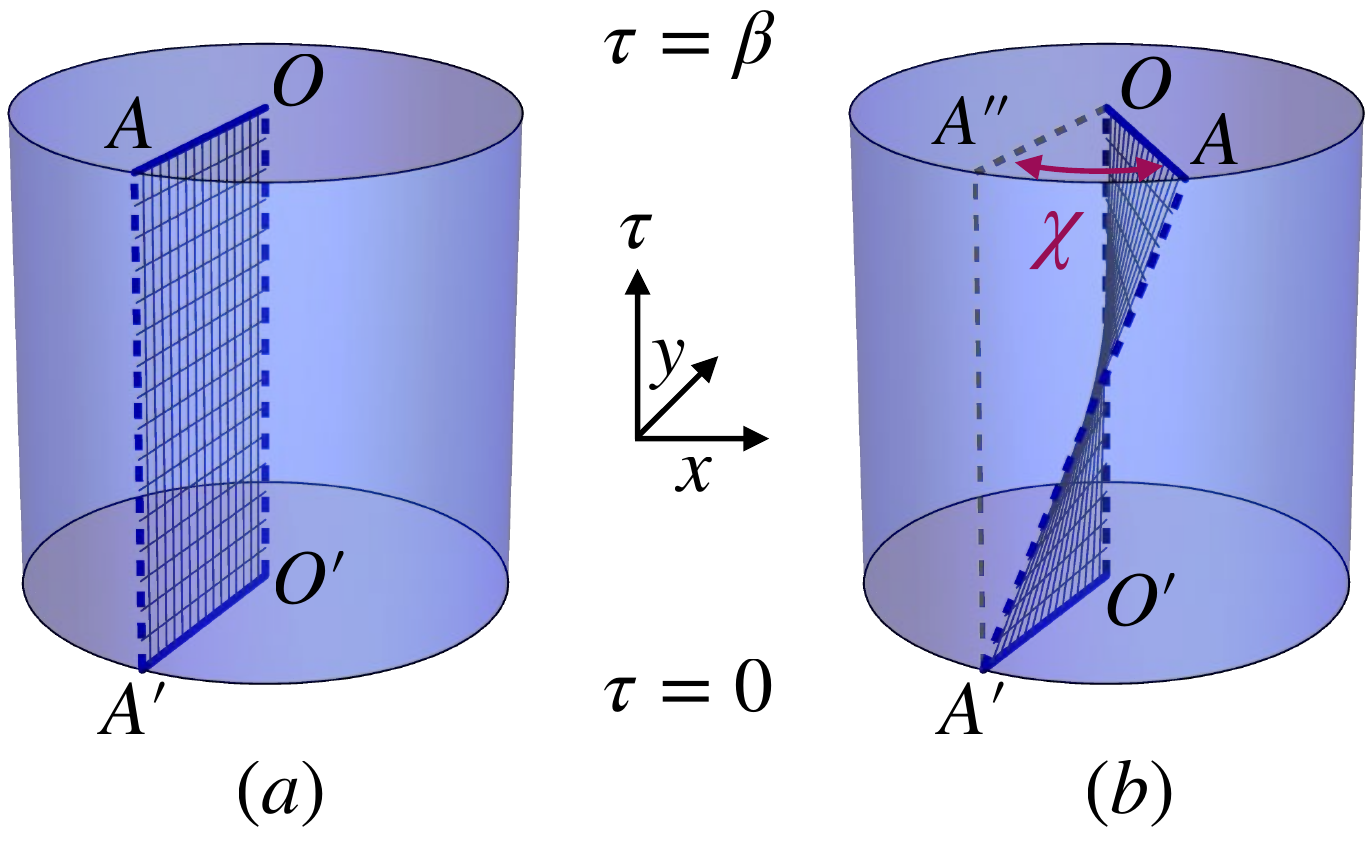}}
\caption{Illustration of (a) standard~\eq{eq_fields_boundary} and (b) rotwisted~\eq{eq_rotwisted} boundary conditions with the segments $O' A'$ at $\tau =0$ and $O A$ at $\tau = \beta$ identified. The axis of rotation $z$ is not shown. The value of the rotwisted angle $\chi$ depends on the imaginary angular frequency~\eq{eq_chi} and plays a role of the statistical parameter.}
\label{fig_cylinder}
\end{figure}

Since the angular frequency, $\Omega$ has the dimension of angle per unit of time, one can also map it, under the Wick transformation, to a purely imaginary quantity:
\beqn
\Omega = i \Omega_I\,.
\label{eq_Wick_Omega}
\eeqn
The imaginary angular frequency $\Omega_I \neq 0$ corresponds to the uniform rotation of the whole spatial timeslice of the Euclidean spacetime~\cite{Chernodub:2020qah}. As the imaginary time variable~$\tau$ advances for a full period from $\tau = 0$ to $\tau = \beta$, the space experiences the rotation by the angle 
\beqn
\chi = \beta \Omega_I\,,
\label{eq_chi}
\eeqn
about the same axis ${\bs n} = {\bs \Omega}_I/\Omega_I$. The imaginary Euclidean rotation modifies the boundary conditions of the fields as illustrated in Fig.~\ref{fig_cylinder}.

The rotation with the imaginary frequency corresponds to the Hamiltonian
\beqn
\hat{\mathcal H}_{{\bs \Omega}_I} = \hat{\mathcal H}_0 - i \hat {\bs \Omega}_I \cdot \hat{\bs {\mathcal J}}\,,
\label{eq_H_Omega_I}
\eeqn
which follows directly from Eqs.~\eq{eq_H_Omega} and \eq{eq_Wick_Omega}. Despite the resemblance of real and imaginary rotations, they correspond to different physical environments~\cite{Chernodub:2020qah,Chen:2022smf}. 

In the absence of rotation, the bosonic (fermionic) wavefunction $\phi$ ($\psi$) is a periodic (anti-periodic) function of the imaginary time $\tau$:
\beqn
\Omega_I = 0: \quad \left\{
\begin{array}{rcl}
\phi({\bs x},\tau) & = & +\, \phi({\bs x}, \tau + \beta)\,, \\[1mm]
\psi({\bs x},\tau) & = & -\, \psi({\bs x}, \tau + \beta)\,.
\end{array}
\right.
\label{eq_fields_boundary}
\eeqn

Under the imaginary rotation, the boundary conditions become as follows:
\beqs
\beqn
\phi({\bs x},\tau) & = & +\, \phi\left({\hat R}_{{\bs \chi}} {\bs x},\tau + \beta\right)\,,
\label{eq_phi_rotation}\\
\psi({\bs x},\tau) & = & -\, {\hat\Lambda}_{\bs \chi} \psi\left({\hat R}_{{\bs \chi}} {\bs x},\tau + \beta\right)\,,
\label{eq_psi_rotation}
\eeqn
\label{eq_rotwisted}
\eeqs
where the $3 \times 3$ matrix ${\hat R}_{{\bs \chi}}$ rotates rigidly the whole spacial Euclidean subspace, ${\bs x} \to {\bs x}' = {\hat R}_{{\bs \chi}} {\bs x}$, by the angle~\eq{eq_chi} ${\bs \chi} = \beta {\bs \Omega}_I$ around the axis ${\bs n} = {\bs \chi}/\chi = {\bs \Omega}_I/\Omega_I$ which corresponds to the axis of the real rotation, as shown in Fig.~\ref{fig_cylinder}. The matrix ${\hat\Lambda}_{\bs \chi}$ represents the rotation in the spinor space. A similar factor should appear for a vector boson. The rotwisted (from ``rotation'' and ``twist'') boundary conditions~\eq{eq_rotwisted} can be implemented in the Euclidean lattice simulations of field theories~\cite{Chen:2022smf,Chernodub:2022veq}.

For definiteness, we consider below the rotation around the $z$ axis, which can be written in the cylindrical coordinates as follows: $(\rho,\varphi,z,\tau) \to (\rho,\varphi - \Omega_I \tau,z,\tau+\beta)$. The boundary conditions~\eq{eq_rotwisted} are visualized in Fig.~\ref{fig_cylinder}. The imaginary rotation in the whole spacetime does not lead to causality problems~\cite{Yamamoto:2013zwa} and can be formulated in the thermodynamic limit in the whole Euclidean space~\cite{Chen:2022smf}. 

The free energy of bosonic and fermionic systems under the imaginary rotation are straightforwardly obtained by identifying the angular frequencies~\eq{eq_Wick_Omega} in the free energy in the co-rotating frame~\eq{eq_F_bos_rot} and \eq{eq_F_ferm_rot}, respectively.

Below we consider free gases of bosonic and fermionic particles subjected to imaginary rotation. The latter is formulated in terms of the boundary condition~\eq{eq_rotwisted} in the imaginary time formalism. In the thermodynamic limit, these gases possess a fractal structure in their thermodynamic properties and host exotic excitations with occupation numbers different from those for bosons and fermions.

\subsection{\bf Fractal thermodynamics of imaginary rotation} 
\label{sec_fractal_thermodynamics}

\subsubsection{Hints from classical bosonic solutions} 

Analysis of classical solutions in Yang-Mills theory has shown that the imaginary rotation with the rational nonzero values of the angular frequency~\eq{eq_chi} corresponds to the thermal bath with uniform temperature~\cite{Chernodub:2022wsw}:
\beqn
T = \frac{1}{q \beta} \qquad \mbox{for} \quad  \frac{\Omega_I \beta}{2 \pi} \equiv \frac{\chi}{2 \pi} = \frac{p}{q}\,,
\label{eq_T_rational_bosons}
\eeqn
where $\beta$ is the length of the lattice in the imaginary-time direction and the rational number $p/q$, with positive integers $p,q = 1,2,\dots$, represents an {\it irreducible fraction}. 

Equation~\eq{eq_T_rational_bosons} provides us with two hints on the behavior of systems under imaginary rotation. First, it suggests that temperature changes in a non-analytical way\footnote{For example, two close values of the imaginary frequency, $\Omega_I \beta/(2 \pi) = 1/2$ and $999/2000$, correspond, respectively, to the temperature values~\eq{eq_T_rational_bosons} $T=1/(2\beta)$ and $T = 1/(2000 \beta)$ that differ from each other by three orders of magnitude.} with the imaginary frequency $\Omega_I$ thus forbidding the analytical continuation from imaginary to real-valued angular frequencies. Second, it stresses the significance of rational numbers which is particular for fractal structures~\cite{hofstadter1976energy}. Thus, Eq.~\eq{eq_T_rational_bosons} provides us with a signature of a fractal behavior of imaginary rotation, which we explore further below. 

\subsubsection{Free bosons at imaginary rotation}
Let us now consider free bosons in thermal equilibrium in the thermodynamic limit. In the cylindrical coordinates, the Hamiltonian for a scalar particle possesses the following eigenfunctions,
\beqn
\phi_{k_\rho,k_z,m}(\rho,z,\varphi) = N_1 e^{i m \varphi} e^{i k_z z} J_m(k_\rho \rho)\,,
\label{eq_wave_function_0}
\eeqn
where $J_m$ is the Bessel function and $N_1$ is a normalization factor. The corresponding energy eigenvalues are
\beqn
\omega_{k_\rho,k_z} = \sqrt{k_\rho^2 + k_z^2 + M^2}\,,
\label{eq_energy_scalar}
\eeqn
where $k_\rho \geqslant 0$ and $k_z \in \R$ are the momenta along the radial direction $\rho$ and the $z$ axis, respectively. The projection of the angular momentum on the $z$ axis, $m \in \Z$, does not enter the expression for the eigenenergy~\eq{eq_energy_scalar} thus corresponding to a degeneracy factor of the eigenstates. The integration measure in Eq.~\eq{eq_F_bos_rot} takes the form:
\beqn
\sumint{\alpha,m} \equiv \int_0^\infty \frac{k_\rho d k_\rho}{2 \pi} \int_{- \infty}^\infty \frac{d k_z}{2\pi} \sum_{m \in \Z}\,. 
\label{eq_measure}
\eeqn

For the rational angular momentum $\Omega_I \beta/(2\pi) = p/q$, we combine the elements of the sum over the angular quantum number $m$ in Eq.~\eq{eq_measure} in the groups of $q$:
\beqn
\sum_{m \in \Z} f(m) = \sum_{\mathfrak{m} \in \Z} \sum_{a=0}^{q - 1} f(q \mathfrak{m} + a)\,, \qquad q = 1,2, \dots \,.
\label{eq_sums_m}
\eeqn
Then the bosonic partition function~\eq{eq_F_bos_rot} at the imaginary angular frequency~\eq{eq_Wick_Omega} can be identically rewritten as
\beqn
F^{\mathrm{(b)}}_\beta & = & \frac{V}{\beta} \int_0^\infty \frac{k_\rho d k_\rho}{2 \pi} \int_{- \infty}^\infty \frac{d k_z}{2\pi} \sum_{\mathfrak{m} \in \Z} \sum_{r = \pm 1} \nonumber \\
& & \hskip 5mm \times \ln \left( 1 - e^{- q \beta (\omega_{k_\rho,k_z} - r \mu)} \right)\!.
\label{eq_F_bos_I}
\eeqn
where we have used the identity (taking $\gamma > 0$):
\beqn
\frac{1}{2} \sum_{c = \pm 1} \sum_{m=0}^{q-1}  \ln \left( 1 - e^{- \gamma + 2 \pi i c m \frac{p}{q}} \right) 
= \ln \left( 1 - e^{- q \gamma} \right)\,.
\label{eq_identity_bosons}
\eeqn
Due to the degeneracy of the energy eigenvalues~\eq{eq_energy_scalar} with respect to the angular momentum~$m$, this identity can be applied to Eq.~\eq{eq_F_bos_I}. Special care is needed to match the infinite sums over $m$ and $\mathfrak m$ which require a regularization. Introducing the ultraviolet regulator in Eq.~\eq{eq_sums_m}, $e^{- \epsilon |m|}$ with $\epsilon > 0$, one finds in the limit $\epsilon \to 0$ that the sums over $m$ and $\mathfrak m$ are not equivalent but differ by the factor of $1/q$, implying $\sum_{m \in \Z} = (1/q) \sum_{\mathfrak{m} \in \Z}$.

Thus, free bosonic gas subjected to the rational imaginary angular frequency~\eq{eq_T_rational_bosons} at temperature $T$ in the absence of background potential has the same free energy as the non-rotating gas at lower temperature $T/q \equiv 1/(q\beta)$:
\beqn
F^{\mathrm{(b)}}_\beta{\biggr|}_{\Omega_I = \frac{2 \pi}{\beta} \frac{p}{q}} = F^{\mathrm{(b)}}_{q \beta}{\biggr|}_{\Omega = 0}\,.
\label{eq_equivalence_bosonic}
\eeqn
This relation complies with Eq.~\eq{eq_T_rational_bosons} obtained for classical bosonic solutions. Notice that Eq.~\eq{eq_equivalence_bosonic} does not depend on the numerator $p$ of the irreducible fraction $p/q$.

The equivalence~\eq{eq_equivalence_bosonic} is easy to understand. The rotation with the angular frequency $\Omega_I = \frac{2 \pi}{\beta} \frac{p}{q}$ rotates the $\bs x$-space by the angle $\chi = 2 \pi \frac{p}{q}$ per one period $\beta$ of the imaginary time $\tau$. After $q$ periods, the spatial $\bs x$-space returns to the original position, thus signaling the $q \beta$ periodicity in the imaginary time\footnote{Here, we do not consider irrational imaginary frequencies $\Omega_I$ for which the periodicity does not exist. Irrational frequencies correspond to a zero temperature according to our discussion below, based on the Thomae function~\eq{eq_Thomae}. The analogy is similar to dynamics of an electron on a square lattice exposed to rational and irrational magnetic fields that have very different effects on the electron dynamics~\cite{hofstadter1976energy}.} . This periodicity length provides us with the temperature of the system, $T = 1/(q \beta)$. While the extended imaginary time contains $q$ copies of original space, the number of particles per one copy remains the same. In short, the rotation of a bosonic thermal gas with an imaginary angular frequency reduces the temperature of the gas by a $q$-factor related to the rational frequency~\eq{eq_T_rational_bosons}.

\subsubsection{Fractal thermodynamics for bosons}

The thermodynamics of the particles under imaginary rotations is curious. In the absence of rotation, $\Omega_I = \Omega = 0$, the density of free energy $f = F/V$ of massless (for simplicity) neutral real-valued scalars at temperature $T = 1/\beta$ is as follows:
\beqn
f^{(b)}_0 \equiv f^{\mathrm{(b)}}_{\beta}{\biggr|}_{\Omega = 0} = - \frac{\pi^2}{90 \beta^4}\,.
\label{eq_F_bosons}
\eeqn
Therefore, for the black-body scalar gas, the energy density $\varepsilon_0 = \partial (\beta f_0)/\partial \beta$, the pressure $P_0 = - f_0$, and the entropy density $s_0 = \beta(\varepsilon_0 + P_0)$, are:
\beqn
\varepsilon_0 = 3 P_0 = \frac{\pi^2}{30 \beta^4}\,,  \qquad  s_0 = \frac{2 \pi^2}{45 \beta^3}\,,
\label{eq_quantities}
\eeqn
where we drop the superscript ``(b)'' for brevity. We also set the chemical potential to zero, $\mu=0$, in the rest of the article. 

The bosonic equivalence~\eq{eq_equivalence_bosonic} implies that the thermodynamic quantities~\eq{eq_quantities} are affected by the rotation with the imaginary angular momentum $\Omega_I$ as follows: 
\beqn
\left(\frac{\varepsilon_{\Omega_I}}{\varepsilon_0}\right)^{\frac{1}{4}} \! 
= \left( \frac{P_{\Omega_I}}{P_0}\right)^{\frac{1}{4}} \! 
= \left(\frac{s_{\Omega_I}}{s_0} \right)^{\frac{1}{3}} \! 
= f_{\mathsf{T}}\left(\frac{\beta\Omega_I}{2 \pi}\right)\,, \quad
\label{eq_quantity_Omega_I}
\eeqn
where
\beqn
f_{\mathsf{T}}(x) =
\left\{
\begin{array}{rl}
    \frac{1}{q} &   \mathrm{if}\ x = \frac{p}{q} \in {\mathbb Q}, \ \mathrm{with}\ p,q\in {\mathbb N} \ \mathrm{coprimes},\\[2mm]
    0 &  \mathrm{if}\  x \notin {\mathbb Q}\,,
\end{array}
\right.
\label{eq_Thomae}
\eeqn
is the Thomae function. It is equal to $1/q$ if its argument is a rational number given by an irreducible fraction, $x = p/q \in {\mathbb Q}$, and zero otherwise. We would like to stress that the scaling behavior of thermodynamic quantities~\eq{eq_quantity_Omega_I} is determined solely by the denominator $q$ given the rational imaginary frequency, as in Eq.~\eq{eq_T_rational_bosons}, and not by its numerator,

\begin{figure}[!thb]
\centerline{\includegraphics[scale=1.07,clip=true]{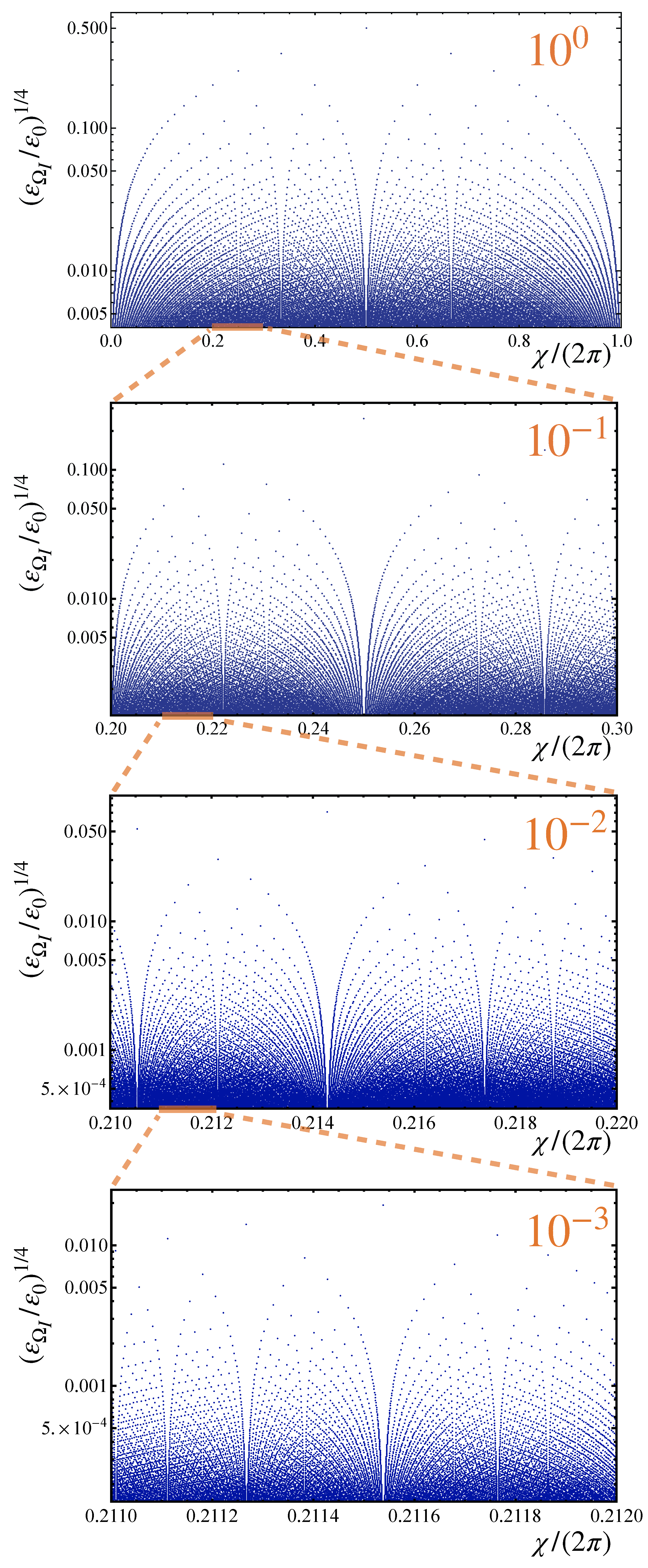}}
\caption{Fractal thermodynamics under imaginary rotation in thermodynamic limit: the characteristics of free massless boson gas as function of the statistical angle $\chi$ related to imaginary angular frequency $\Omega_I$ in Eq.~\eq{eq_chi}. We show only the energy density as it is connected to other thermodynamic quantities via Eq.~\eq{eq_quantity_Omega_I}. The normalization to the non-rotating gas is implied~\eq{eq_quantities}. The plots show various zoom scales from $10^0$ to $10^{-3}$ which show the self-similar pattern generated by the Thomae function~\eq{eq_Thomae}.}
\label{fig_fractal_energy}
\end{figure}

In Fig.~\ref{fig_fractal_energy}, we show the energy density of a massless boson gas as a function of the statistical angle $\chi$, which is related to the imaginary angular velocity~$\Omega_I$ in Eq.~\eq{eq_chi}. According to Eq.~\eq{eq_quantity_Omega_I}, this thermodynamic quantity is normalized to the non-rotating boson gas value at a fixed temperature~\eq{eq_quantities}. Thermodynamics exhibits an explicit fractal behavior characterized by a self-repeating structure at all temperature scales. If plotted on the logarithmic scale, the visual appearance of thermodynamic quantities resembles (fractal) water fountains. The self-similarity patterns in this fractal structure, generated by the Thomae function~\eq{eq_Thomae}, are given by the denominators of the successive Farey sequences of fractions~\cite{Devaney1999}.

We have also found the fractal-like behavior of free-fermion thermodynamics, similar to the bosonic fractal fountains shown in Fig.~\ref{fig_fractal_energy}.

\subsection{A no-go theorem for analytical continuation}
\label{sec_no_go}

The absence of analytical continuation is evident from the non-analytical nature of the fractal behavior of thermodynamics~\eq{eq_quantity_Omega_I} at imaginary angular frequencies~$\Omega_I$. The fractal is generated by the Thomae function~\eq{eq_Thomae} which is discontinuous at every rational point because irrational numbers come infinitely close to any given rational number. Thus, at every value of $\Omega_I$, the thermodynamic quantities are non-differential, non-analytical functions of the imaginary angular frequency $\Omega_I$.

\begin{figure}[!thb]
\centerline{\includegraphics[scale=0.45,clip=true]{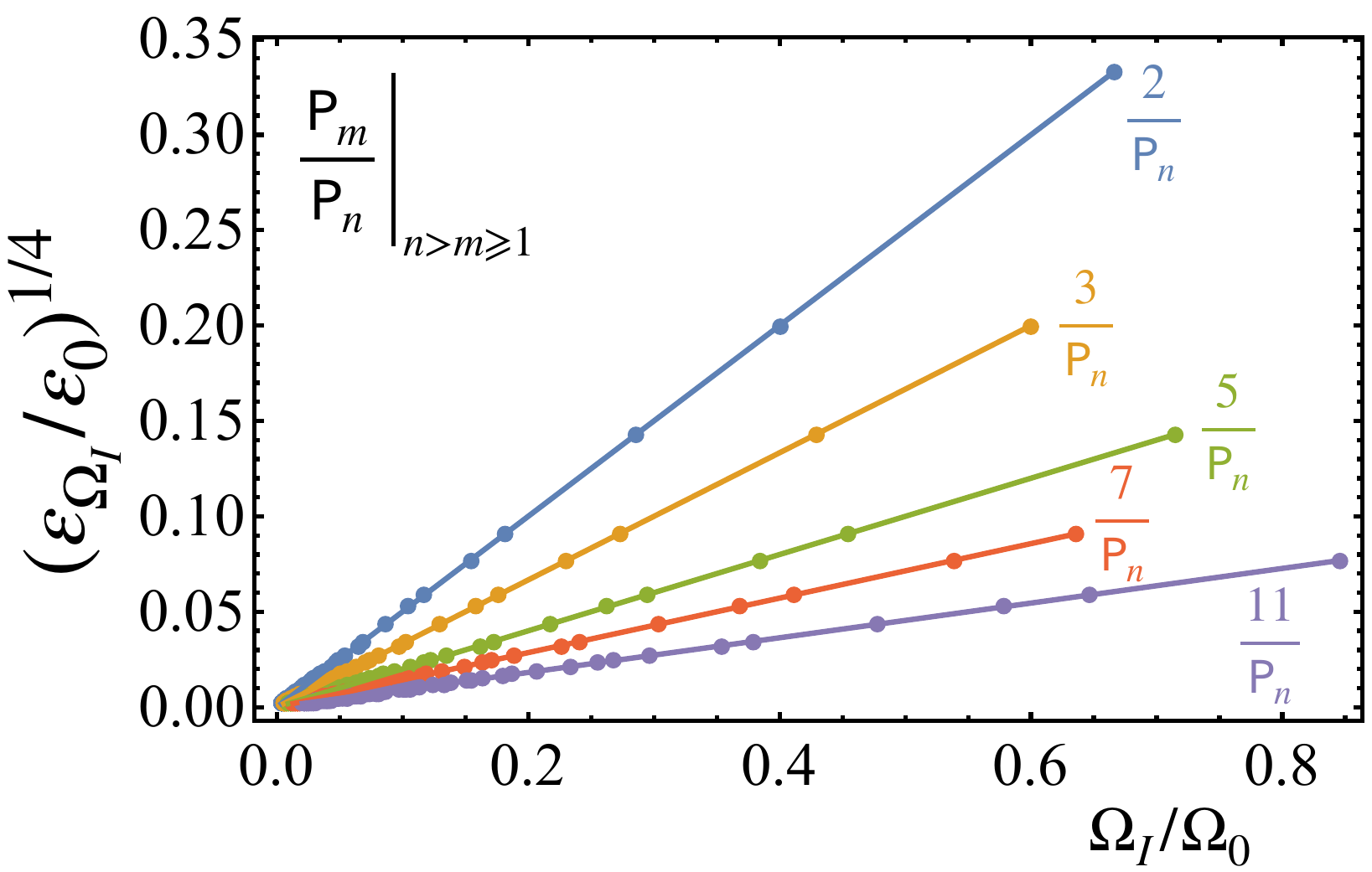}}
\caption{Absence of analytical continuation from imaginary to real rotation for free bosonic system: the energy density vs. the imaginary angular frequency for different prime-number sequences, $\Omega_I = 2 \pi \beta^{-1} {\mathsf P}_m/{\mathsf P}_n$ in units of $\Omega_0 = 2 \pi T$.}
\label{fig_analytical}
\end{figure}

The no-go statement is illustrated in Fig.~\ref{fig_analytical} where the energy density is shown as the function of the imaginary angular momentum $\Omega_I$ for various sequences $p/q = {\mathsf P}_m/{\mathsf P}_n$, with $m>n$, of the prime numbers ${\mathsf P}_i$. The slopes of these sequences are different, thus demonstrating the ambiguity of the analytical continuation (or invalidity of any analytical method based, for example, on a Taylor expansion). Since the fractal properties of imaginary rotations are common for bosons and fermions, the same no-go argument is also valid for both types of particles.

In concluding this section, we would like to add four comments. First of all, we notice that although our no-go arguments are based on the behavior of free theories and the properties of classical solutions, they should also work in interacting models (at least, in perturbation theory over classical backgrounds). 

Second, the analytical continuation can exist for non-local quantities. For example, the Polyakov loop in gauge field theories~\cite{Chernodub:2022veq} allows us to formulate a Euclidean analog of the Tolman-Ehrenfest law in rotating systems. The Minkowski and Euclidean Tolman-Ehrenfest laws appear to be connected by an analytical continuation between real and imaginary angular frequencies. One can show that this analytical continuation has a pure kinematic origin.

Third, the absence of analytical continuation between real and imaginary rotations for local thermodynamic quantities implies that the imaginary rotation is principally different and disconnected from the rotation with real-valued angular momentum. This conclusion is valid in the thermodynamic limit. Below, we show that the imaginary rotation leads to a new type of particle statistics.

Fourth, our conclusions are valid in the strict thermodynamic limit where the system under the imaginary rotation is considered in the infinite volume.

\section{Statistical transmutation}
\label{sec_transmutation}

\subsection{Examples of statistic transmutations}
\label{sec_examples_transmutation}

\subsubsection{Transmutation of fermions to bosonic ghosts} 
\label{sec_trans_ferm_bos}
To illustrate the concept of our idea, let us start with a few examples. In Ref.~\cite{Chernodub:2022wsw}, we have shown that the rotation with the imaginary frequency $\Omega_I = 2 \pi/\beta$ converts fermions into ghost-like particles that behave as bosons but contribute to the free energy with the wrong, for bosons, sign. Here, we explore this property further for a generic rational frequency by repeating the above arguments for fermionic particles. We skip most of the derivation, which is trivial. Instead of the bosonic identity~\eq{eq_identity_bosons}, we use the following relation for fermions:
\beqn
& & \frac{1}{2} \sum_{c = \pm 1} \sum_{m=0}^{q-1}  \ln \left( 1 + e^{- \gamma + 2 \pi i c (m + \frac{1}{2}) \frac{p}{q}} \right) 
\nonumber \\
& & \hskip 14mm = \ln \left[1 - (-1)^{p + q} e^{- q \gamma} \right]\,,
\label{eq_identity_fermions}
\eeqn
which leads us to the identities:
\beqn
F^{\mathrm{(f)}}_\beta {\biggr|}_{\Omega_I = \frac{2 \pi}{\beta} \frac{p}{q}} =
\left\{
\begin{array}{rcl}
F^{\mathrm{(f)}}_{q \beta}{\Bigr|}_{\Omega = 0}, & & p+q \in \mathrm{odd}, \\[3mm]
- 2 F^{\mathrm{(b)}}_{q \beta}{\Bigr|}_{\Omega = 0}, & & p+q \in \mathrm{even}.
\end{array}
\right.
\label{eq_equivalence_fermionic}
\eeqn
The effect of the imaginary angular frequency on temperature and degeneracy is exactly the same as for the bosons: an imaginary-rotated fermion ensemble gets temperature $T = 1/(q\beta)$, as in Eq.~\eq{eq_T_rational_bosons}.

The mapping~\eq{eq_equivalence_fermionic} shows that for odd numbers $p+q$, the rotating gas is described by fermions, as expected. However, for even $p+q$, the rotation produces ghosts out of fermions. The ghosts obey bosonic statistics but have free energy with the wrong (fermionic) sign:
\beqn
F^{\mathrm{(gh)}}_{\beta} \equiv - F^{\mathrm{(b)}}_{\beta}\,. 
\label{eq_ghost_boson}
\eeqn
The factor 2 in Eq.~\eq{eq_equivalence_fermionic} corresponds to a spin degeneracy: under the imaginary rotation, one Dirac fermion produces two ghosts~\eq{eq_equivalence_fermionic}. 

The thermodynamics of fermions under the imaginary rotation also has the ``fractal fountain'' structure similar to the one of Fig.~\ref{fig_fractal_energy}. The fractal fountains, however, acquire a two-color pattern due to even/odd partitioning of the free energy~\eq{eq_equivalence_fermionic}. The relations~\eq{eq_equivalence_fermionic} can be understood along the same analysis that we performed in the case of bosons, taking into account the anti-periodicity of fermions in the $\tau$ direction. 

For fermions, the rotation with the imaginary frequency $\Omega_I = 2 \pi/\beta$ corresponds to an exceptional case due to anti-periodicity of fermionic fields. Therefore, for fermions, the irreducibility of $p/q$ is allowed for $p = q = 1$. A full rotation of the system at the angle $\chi = 2 \pi$ after one period changes the sign of the fermionic wavefunction, which, together with the original anti-periodicity~\eq{eq_psi_rotation}, gives the periodic boundary conditions for the fermionic fields at temperature $T = 1/\beta$. Thus, the imaginary rotation at the frequency $\Omega_I = 2\pi/\beta$ transmutes fermions into bosonic ghosts that have bosonic statistics but enter the free energy with the wrong, ``fermionic'' sign.

\subsubsection{Transmutation of bosons to fermionic ghosts} 
\label{sec_examples_transmutation_bosons}

\begin{figure}[!thb]
\centerline{\includegraphics[scale=0.45,clip=true]{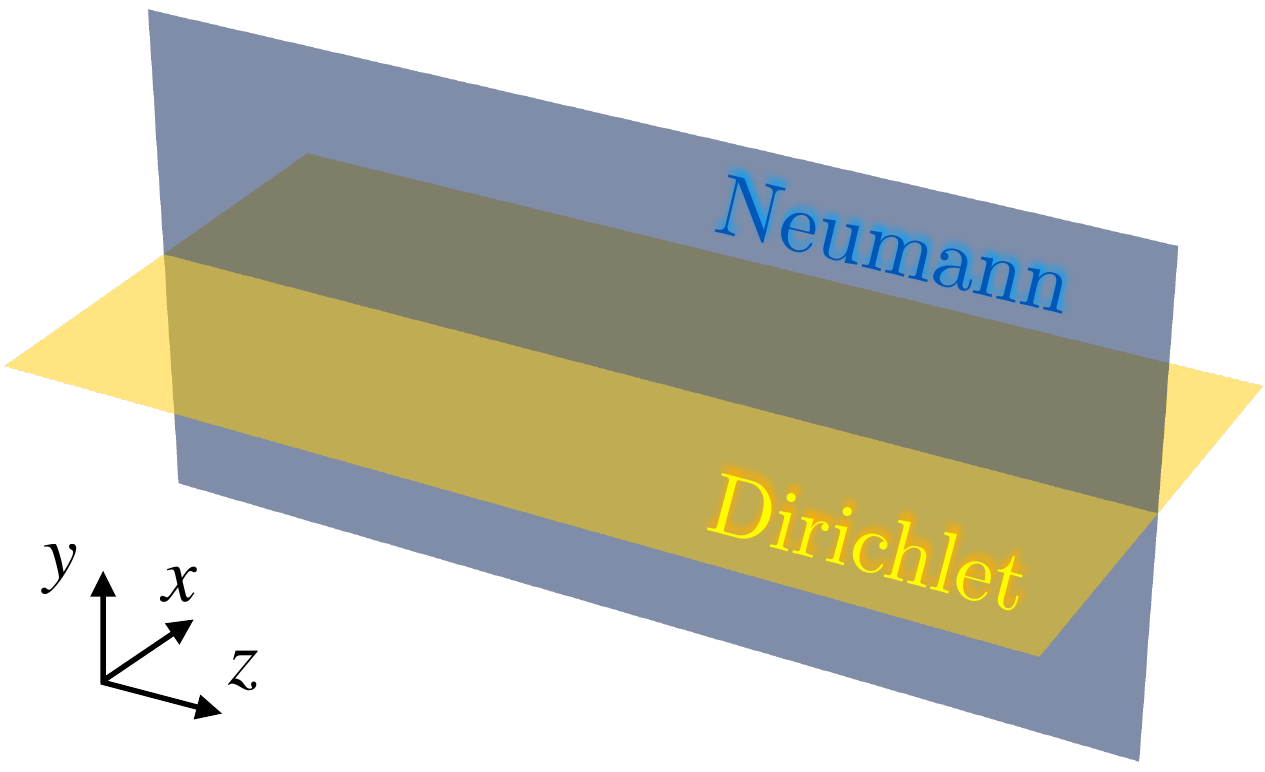}}
\caption{Intersecting Dirichlet and Neumann walls~\eq{eq_bc_ghost}. The potential set by the walls is consistent with the imaginary-time boundary conditions~\eq{eq_phi_rotation} set by the imaginary rotation $\Omega_I = \pi/\beta$ (the statistical parameter $\chi = \pi$). Subjected to imaginary rotation, this system converts a boson into a fermionic ghost with the free energy~\eq{eq_F_bos_rot_pi} and thermodynamic characteristics given in Eqs.~\eq{eq_cross_thermodynamics_1}.}
\label{fig_walls}
\end{figure}

The ghosts can emerge not only in fermionic~\eq{eq_ghost_boson} but also in bosonic systems. Let us consider, for example, a massless boson $\phi$ in the space with two crossing infinitely large walls as shown in Fig.~\ref{fig_walls}. The walls are oriented in the $xz$ and $yz$ planes and intersect along the $z$ axis. We impose that these $xz$ and $yz$ walls set the Dirichlet and Neumann (DN) boundary conditions, respectively:
\beqn
{\mathrm{DN}}: \quad \phi({\bs x},\tau){\biggr|}_{y = 0} = 0, 
\qquad \frac{\partial \phi({\bs x},\tau)}{\partial x}{\biggr|}_{x = 0} = 0\,,
\label{eq_bc_ghost}
\eeqn

In the cylindrical system of coordinates ($\rho,\varphi,z$), the wavefunctions consistent with the boundary conditions~\eq{eq_bc_ghost} are as follows:
\beqn
\phi_{k_\rho,k_z,m}(\rho,z,\varphi) = N_2 \sin(m \varphi) e^{i k_z z} J_m(k_\rho \rho)\,.
\label{eq_wave_function_1}
\eeqn
where $N_2$ is a normalization factor. The energy spectrum takes the form~\eq{eq_energy_scalar} with one important modification: the walls diminish the level of the degeneracy of the states by restricting the angular momentum to odd positive numbers: $m = 2 {\mathfrak m} + 1$ with ${\mathfrak m} = 0, 1, \dots$. This restriction leads to a factor of ``$1/4$'' in thermodynamic quantities. In the absence of rotation, the energy density, pressure, and entropy of the crossed wall system of Fig.~\ref{fig_walls} (denoted by ``\ding{59}'') can be derived from the generic expression for the bosonic free energy~\eq{eq_F_bos_rot}.
One gets:
\beqn
\varepsilon_{\Omega_I = 0}^{(\littlecross)} = 3 P_{\Omega_I = 0}^{(\littlecross)} = \frac{\pi^2}{120 \beta^4}\,, \qquad 
s_{\Omega_I = 0}^{(\littlecross)} = \frac{\pi^2}{90 \beta^3}\,.
\label{eq_cross_thermodynamics_0}
\eeqn
These quantities are, expectedly, four times smaller than their counterparts in the system without walls~\eq{eq_quantities}.

Consider now the imaginary rotation about the axis $z$ with the imaginary angular frequency $\Omega_I = \pi/\beta$. Setting $\Omega = i \pi/\beta$ in the expression for the free energy~\eq{eq_F_bos_rot}, one gets  $e^{\beta c m \Omega} = - 1$ for odd angular momentum $m$. This sign flip changes (``transmutes'') effectively the statistics in the bosonic free energy~\eq{eq_F_bos_rot}:
\beqn
F^{\mathrm{(b)}}_\beta{\biggl{|}_{\Omega_I = \frac{\pi}{\beta}}} = \frac{2 V}{\beta} \sumint{\alpha} \sum_{{\mathfrak{m}}=0}^\infty \ln \left( 1 + e^{- \beta \omega_{\alpha,{2 {\mathfrak{m}} + 1}}} \right)\!.
\label{eq_F_bos_rot_pi}
\eeqn
The free energy of the imaginary rotating system~\eq{eq_F_bos_rot_pi} becomes similar to the free energy of fermions~\eq{eq_F_ferm_rot} but with the wrong, ``bosonic'' sign. Thus, according to our definition above~\eq{eq_ghost_boson}, the bosons have been transmuted into fermionic ghosts.

Apart from the transmutation of statistics, the rotation with the imaginary angular frequency $\Omega_I = \pi/\beta$ leads, according to Eq.~\eq{eq_T_rational_bosons}, to the decrease of temperature by the factor of 2 with $\beta \to 2 \beta$. Finally, taking into account the mentioned reduction factor 1/4 due to the presence of the walls, we obtain that the bosonic gas under the cross potential of Fig.~\eq{fig_walls} subjected to the imaginary rotation with the frequency $\Omega_I = \pi/\beta$ becomes the gas of fermionic ghosts characterized by the exotic thermodynamics:
\beqn
\varepsilon_{\Omega_I = \frac{\pi}{\beta}}^{(\littlecross)} {=} 3 P_{\Omega_I = \frac{\pi}{\beta}}^{(\littlecross)} {=} - \frac{\pi^2}{1920 \beta^4}\,, \quad s_{\Omega_I = \frac{\pi}{\beta}}^{(\littlecross)} {=} - \frac{\pi^2}{1440 \beta^3}. \  \quad
\label{eq_cross_thermodynamics_1}
\eeqn

Notice that for the statistical transmutation of free bosons to the fermionic ghosts, one needs the presence of both a background potential and an imaginary rotation. The previous subsection shows that free fermions can be transmuted into bosonic ghosts only under an imaginary rotation without a background potential.

\subsubsection{Thermodynamics of ghosts: examples} 

According to relations~\eq{eq_F_bosons}, \eq{eq_quantities}, \eq{eq_equivalence_fermionic} and \eq{eq_ghost_boson}, the ghosts possess rather exotic thermodynamic features characterized by negative energy density, negative pressure, and negative entropy. These quantities should decrease with the temperature increase (while being negative, they increase in absolute value). 

The unusual properties of the thermodynamic ensemble, generated by the rotwisted boundary conditions~\eq{eq_rotwisted}, can be probed in first-principle Monte Carlo calculations. One can mention simulations of Yang-Mills theory at the imaginary angular momentum $\Omega_I = \frac{1}{4} \times 2 \pi/\beta$ which is consistent with symmetries of the hypercubic lattice~\cite{Chernodub:2022veq}. The theoretical considerations on the same imaginary frequency have also been presented in Ref.~\cite{Chen:2022smf}.

In principle, the lattice field theory can also be formulated at a triangular lattice which also admits the value of the imaginary angular frequency $\Omega_I = \frac{1}{3} \times 2 \pi/\beta$. According to Eq.~\eq{eq_equivalence_fermionic}, such imaginary rotation leads to the transmutation of one free Dirac fermion into two bosonic ghosts at temperature $T=1/(3 \beta)$ with the energy density, pressure and entropy becoming negative quantities:
\beqn
\Omega_I = \frac{2 \pi}{3\beta} \to
\left\{ 
\begin{array}{l}
\varepsilon_{\Omega_I}^{\mathrm{(gh)}} {=} 3 P_{\Omega_I} = - \frac{\pi^2}{1215 \beta^4}\,, \\[2mm]
s_{\Omega_I}^{\mathrm{(gh)}} {=} - \frac{4 \pi^2}{1215 \beta^3}\,.
\end{array}
\right.
\label{eq_ghost_thermodynamics}
\eeqn
The meaning of negative energy and entropy remains to be understood. One of the interpretations is that these ghosts could be interpreted as collective excitations in a many-body system. While, formally, the entropy of specific excitations can take negative values, the whole many-body system should possess a positive entropy. 

These bosonic and fermionic ghosts can also play a distinctive role as a candidate of the cosmological dark energy used to explain the observed acceleration of the expansion of the Universe~\cite{Peebles:2002gy}.

The negative values of thermodynamic quantities~\eq{eq_ghost_thermodynamics} can be interpreted as a signature of the object to carry a negative number of degrees of freedom. Therefore, we employ the term ``ghost'' in our paper. The same terminology has been used in a different context but with a similar purpose. These are the Faddeev-Popov ghosts that appear in a gauge-fixing procedure in gauge theories~\cite{Faddeev:1967fc} and are utilized to subtract the unphysical degrees of freedom from the gauge bosons. 

The negative values of the thermodynamic quantities for the ghost particles~\eq{eq_ghost_thermodynamics} do not imply the emergence of any (tachyonic) instability in the system. For example, in systems with a finite spatial size, the vacuum pressure can take negative values while the system is perfectly stable. This phenomenon is associated with the Casimir effect~\cite{ref_Casimir} which appears due to the presence of physical objects that affect fluctuations of quantum fields around them. The latter effect changes both energy and pressure of zero-point vacuum fluctuations~\cite{ref_Bogdag, ref_Milton} and produces experimentally observable forces~\cite{link_experiment_1}.

Our ghosts have the same effect on the vacuum energy as the Casimir phenomenon. At particular statistical parameter values, the ghosts can make energy density and pressure negative. Moreover, the ghosts also contribute negatively to the entropy of the system~\eq{eq_ghost_thermodynamics}. This negative entropy phenomenon has also been noticed to appear in certain Casimir systems~\cite{Balian:1977qr,Milton:2017ghh,Bordag:2018lfb,Milton:2018zpc}.

\subsection{Level-dependent statistics under imaginary rotation} 

\subsubsection{Ninion thermodynamics}

The fermionic and bosonic ghosts represent particular cases of a more general excitation which we call a ``ninion''\footnote{
\label{footnote}
From  ``{\it neither} bosons {\it nor} fermions'', with the use of the negative conjunction construction ``{\it ni} \dots {\it ni} \dots'' ({\it neither} \dots {\it nor} \dots) which exists in French and certain Slavic languages.}. Similarly to the ghost particle, the signature of a ninion statistics can be seen in its thermodynamic contribution to the free energy of the system. 

Let us consider a boson or a fermion with free energy, Eq.~\eq{eq_F_bos_rot} or~\eq{eq_F_ferm_rot}, at the imaginary angular frequency~\eq{eq_Wick_Omega}. We subject the particle to a background potential $V({\bs x})$, which (partially) lifts the degeneracy of the energy spectrum on the angular momentum $m$. We do not specify the form of the potential but require that it should 
\begin{itemize}
    \item[(i)] be consistent with the rotwisted boundary conditions~\eq{eq_rotwisted} corresponding to the imaginary frequency~$\Omega_I$, and
    \item[(ii)] admit the quantization of the wavefunctions in terms of the angular quantum number $m$.\footnote{The latter requirement does not imply the invariance of the potential $V({\bs x})$ under a group of continuous rotations about the $z$ axis. An example is given by the non-axially-symmetric background~\eq{eq_bc_ghost}, visualized in Fig.~\ref{fig_walls}, which possess eigenfunctions~\eq{eq_wave_function_1} labeled by the angular momentum~$m$.}
\end{itemize}

The corresponding free energy~\eq{eq_F_bos_rot} takes the following form:
\beqn
F^{\mathrm{(a)}}_\beta & = & \pm \frac{V}{2 \beta} \sumint{\alpha,m} \!\! \sum_{r = \pm 1}\, \ln \bigl( 1 \mp 2 e^{- \beta (\omega_{\alpha,m} {-} r \mu)} \cos \xi^{\mathrm{(a)}}_m \nonumber \\
& & \hskip 15mm +  e^{- 2 \beta (\omega_{\alpha,m} - r \mu)} \bigr), \quad {\mathrm a} = {\mathrm{b,f}},
\label{eq_F_bos_rot_2}
\eeqn
where upper (lower) sign corresponds to bosons, $\mathrm a = b$ (fermions, $\mathrm a = f$). The auxiliary statistical parameters,
\beqn
\xi^{\mathrm{(b)}}_m = m \chi\,, 
\qquad 
\xi^{\mathrm{(f)}}_m = \left( m + \frac{1}{2} \right)\chi\,,
\label{eq_chi_bos_ferm}
\eeqn
depend on the angle $\chi \equiv \beta \Omega_I \in (-\pi,\pi]$ set by the rotwisted boundary condition~\eq{eq_chi}. We remind that the collective index $\alpha$ in Eq.~\eq{eq_F_bos_rot_2} includes all labels of the quantum states except for the angular momentum~$m$ and that the Dirac fermionic index $\alpha$ also contains a sum over the spin projection $s_z = \pm 1/2$. 

In free space, the particle energy does not depend on the angular momentum $m$. Therefore, all levels sum up in the free energy~\eq{eq_F_bos_rot_2}, which reduces, via the identities~\eq{eq_identity_bosons} and \eq{eq_identity_fermions}, to its bosonic~\eq{eq_equivalence_bosonic} and fermionic~\eq{eq_equivalence_fermionic} representations, that exhibit fractal thermodynamics. If the particles interact with a background potential, the contributions of different levels at a nonvanishing statistical angle $\chi \neq 0$ become distinguishable.

The total particle density,
\beqn
N^{\mathrm{(a)}} = - \frac{1}{\beta} \frac{\partial F^{\mathrm{(a)}}}{\partial \mu} \equiv \sumint{\alpha,m} n^{\mathrm{(a)}}_{\omega_{\alpha,m}}(\xi^{\mathrm{(b,f)}}_m), \quad {\mathrm a} = {\mathrm{b,f}},\quad
\eeqn
is determined by the sum over the occupation numbers over energy levels characterized by the angular momentum $m$ and the collective quantum number $\alpha$. Using the expression for the free energy~\eq{eq_F_bos_rot_2} with the rotwisted boundary conditions, we get the following expression for the ``ninionic'' occupation numbers:
\beqn
n^{\mathrm{(a)}}_\omega (\xi) = \frac{e^{\beta (\omega - \mu)}  \cos \xi \mp 1}{1 \mp 2 e^{\beta (\omega - \mu)} \cos \xi + e^{2 \beta (\omega - \mu)}}, \quad {\mathrm a} = {\mathrm{b,f}}.\quad
\label{eq_n_chi}
\eeqn
The upper and lower signs indicate that the ninion statistics originate from bosonic ($\mathrm a = \mathrm b$) and fermionic ($\mathrm a = \mathrm f$) particles, respectively. The continuous parameter~$\xi$, defined for bosons and fermions in Eq.~\eq{eq_chi_bos_ferm}, controls the population of states and interpolates between different kinds of statistics. For example, using the notation $\varepsilon = \beta(\omega - \mu)$, for brevity:
\begin{itemize}

    \item at the angle $\xi=0$, the population numbers~\eq{eq_n_chi} correspond to standard bosonic, $n^{\mathrm{(b)}} = 1/(e^\varepsilon - 1)$, and fermionic, $n^{\mathrm{(f)}} = 1/(e^\varepsilon + 1)$, distributions.

    \item at the angle $\xi=\pi/2$, the bosons become fermionic ghosts $n^{\mathrm{(b)}} = - 1/(e^{2\varepsilon} + 1)$ while the fermions remain fermions $n^{\mathrm{(f)}} = 1/(e^{2\varepsilon} + 1)$, albeit, in both cases, at twice lower temperature, $T = 1/(2 \beta)$;

    \item at the angle $\xi=\pi$, the bosons become fermionic ghosts $n^{\mathrm{(b)}} = - 1/(e^{\varepsilon} + 1)$ and the fermions become bosonic ghosts $n^{\mathrm{(f)}} = - 1/(e^{\varepsilon} - 1)$ at $T = 1/\beta$.
\end{itemize}

\begin{figure}[!thb]
\centerline{\includegraphics[scale=0.6,clip=true]{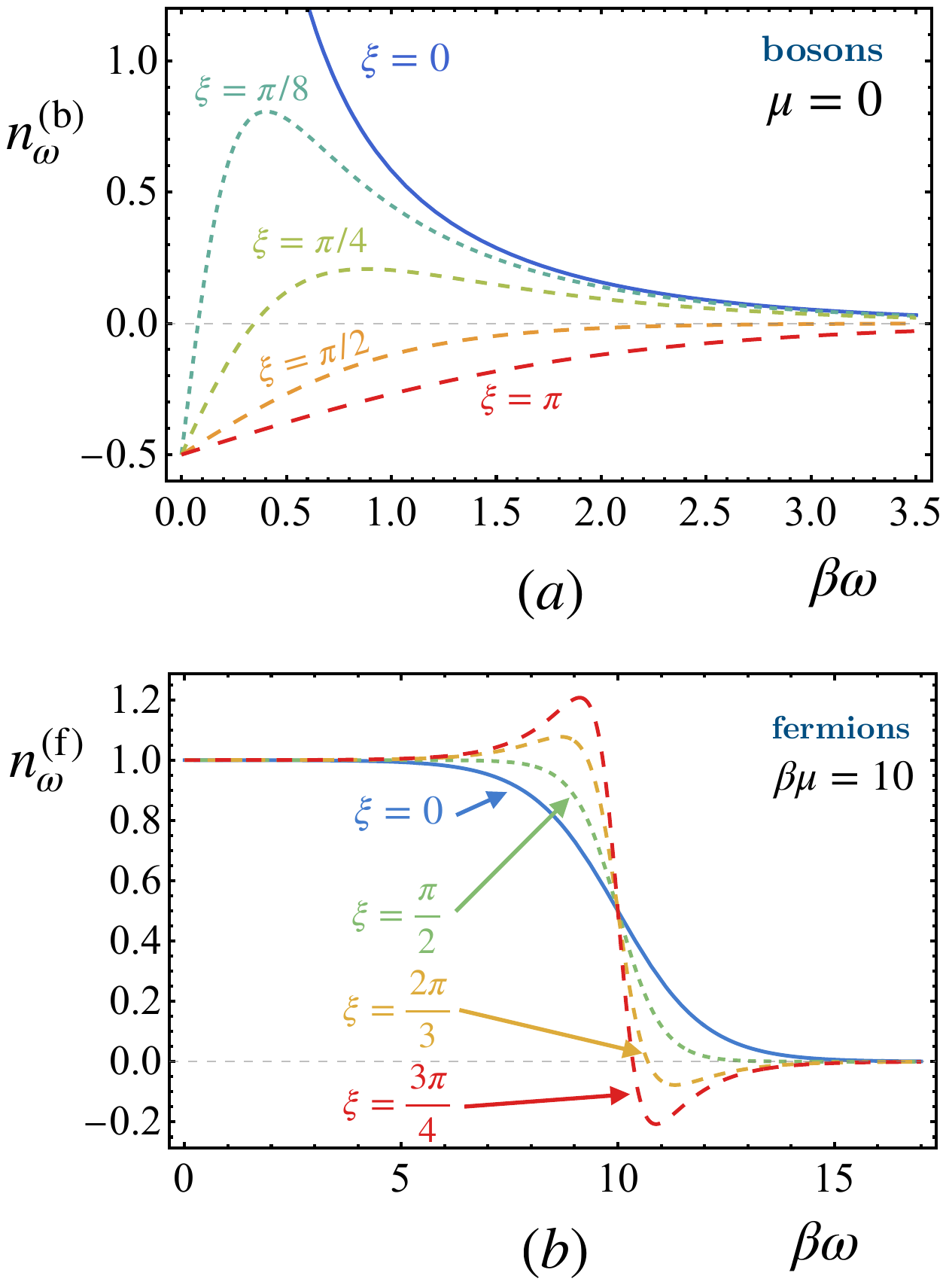}}
\caption{The examples of the ninionic~\eq{eq_n_chi} thermal occupation numbers corresponding to (a) bosons and (b) fermions in the thermal equilibrium for a set of the parameters~$\xi$. The latter are related to the imaginary angular frequency and the statistical parameter $\chi$ via Eqs.~\eq{eq_chi} and \eq{eq_chi_bos_ferm}. The case $\xi = 0$ corresponds to the standard Bose-Einstein and Fermi-Dirac distributions, respectively.}
\label{fig_distributions}
\end{figure}

The examples of the particle occupation numbers~\eq{eq_n_chi} are shown in Fig.~\ref{fig_distributions}.

The occupation numbers~\eq{eq_n_chi} can take negative values. For example, the ninionic occupation number, which originates from bosons, tends to the universal value in the high-temperature limit:
\beqn
\lim_{\beta \to 0} n^{\mathrm{(b)}}_\omega(\xi) = - \frac{1}{2}\,, \qquad \xi \neq 0 \ \mathrm{mod} \ 2\pi\,.
\eeqn
This property is seen in Fig.~\ref{fig_distributions}(a). On the other hand, under the imaginary rotation, the fermions give rise to the ninions that have a negative occupation number just above the Fermi surface. This effect is visible in Fig.~\ref{fig_distributions}(b). 

We suggest that ninions can appear in the many-body systems where the negative occupation number from a ninion is accompanied by a larger positive contribution from other states so that the total particle density remains non-negative. The physical circumstances of the negative density of ninions should be similar to the physics of holes that live below Fermi surfaces in electronic systems. In the latter case, removing an electron from a fermionic sea does not lead to the ``negative density of electrons''. Instead, this process creates a positive density of the holes. In other words, the absence of a negatively charged electron can be understood as the presence of a positively charged hole. In the particle physics context, a similar analogy is given by Dirac's positron, which can be associated with an electron removed from the Dirac sea. We expect that the same idea also applies to the ninions.

\subsubsection{Ninions, anyons, and spin-statistics}

For a non-trivial value of parameter $\xi$, the ninionic statistical distribution~\eq{eq_n_chi} is different from occupation numbers of bosons and fermions. Since the thermal distribution of particle states is inherently connected to the type of statistics of the particle, one could attempt to associate the ninions with particles possessing unconventional statistics. However, this attempt faces a contradiction with the spin-statistics theorem in $3+1$ dimensions.

Indeed, the existence of ninions is a generic property of any system that admits spatial rotations; therefore, the ninions can emerge in $d \geqslant 2$ spatial dimensions. In particular, the ninion can exist in $d=3$ spatial dimensions where the spin-statistics theorem forbids any statistics different from bosonic or fermionic~\cite{Finkelstein:1968hy}. Therefore, let us reiterate that the ninion cannot be associated with a particle possessing a {\it definite} statistics different from the one of bosonic or fermionic particles.

Our results demonstrate that the ninionic property is associated with the energy level of the system rather than with the particle itself. While the occupation number of a ninion~\eq{eq_n_chi} carries a trace of information about the original statistics of this particle, the form of the statistical distribution depends on the quantum level occupied by the particle. For example, a fermionic particle can be a ghost occupation number at one quantum level, and it can again acquire a fermionic occupation number at the other level. 

To illustrate this property, let us consider the occupation number~\eq{eq_n_chi} for a bosonic particle at the statistical parameter $\chi = \pi$. Let us subject the particle to a potential that lifts the degeneracy of the energy levels with respect to the angular momentum (a remaining degeneracy of reflections, $m \to - m$, does not play any role). Then, according to our discussion after Eq.~\eq{eq_n_chi}, the thermal occupation number corresponds to
\begin{itemize}
\item[(i)] a boson at temperature $T = 1/\beta$ for $m=4k$; 
\item[(ii)] a fermionic ghost at $T = 1/(2\beta)$ for $m = 2 k + 1$, 
\item[(iii)] a fermionic ghost at $T = 1/\beta$ for $m = 4 k + 2$, 
\end{itemize}
with $k \in {\mathbb Z}$. The form of thermal distribution depends on the energy level occupied by the particle. 

The non-trivial fractional statistics is often attributed to anyons which are particle-like objects associated with charge-flux composites~\cite{Wilczek:1981du} that also have various possible realizations in solid-state systems (for a recent review, see Ref.~\cite{Greiter:2022iph}). However, the anyons are different from the ninions due to the following three reasons:
\begin{enumerate}
    \item  The anyons, as locally defined fields, exist only in two spatial dimensions since the spin-statistics theorem does not apply to (2+1)d, and particles can be of any spin. The ninions, on the contrary, exist in $d \geqslant 2$ spatial dimensions. 
    
    \item An anyon is a local object that remains the anyon at any quantum level of the system. However, the statistics of ninions --seen via the thermal distributions-- depends on the level of the system.

    \item The commutation relations of a particle field reflect its statistics. For an anyon, the commutation relation differs from its bosonic and fermionic counterparts. In short, for an anyonic field $c$, one has $c(x) c(x') = e^{i \theta} c(x') c(x)$ with $\theta \in (-\pi, \pi]$. However, as we show below, the commutation relation for the ninion fields corresponds to conventional bosonic or fermionic relations, thus excluding an identification of a ninion with an anyon.
\end{enumerate}

\section{Coherent rotational states}
\label{sec_coherent}

\subsection{Rotwisted boundary in path-integral formalism}

In the standard approach to a field theory in thermal equilibrium, bosonic and fermionic fields should satisfy the periodic and anti-periodic boundary conditions in the imaginary time direction. What is the thermodynamic sense behind the rotwisted boundary conditions~\eq{eq_rotwisted} that differ from both mentioned boundary conditions? Here we follow Ref.~\cite{Giusti:2010bb} where the same question has been addressed with the translationally shifted boundaries.

Consider the grand canonical ensemble of a fi\-ni\-te-tem\-perature finite-density spin-0 bosonic system described by the Hamiltonian $\hat{\mathcal H}$ with the particle number operator $\hat{\mathcal N}$. The relative contribution to the partition function ${\mathcal Z}$ coming from the states with the definite projection of the total angular momentum $m$ on a fixed axis ${\bs n}$ is given by the following expectation value:
\beqn
R_{\bs n}\bigl(\beta,\mu; m\bigr) = \avr{\hat{\mathrm P}^{(m)}_{\bs n}} = 
\frac{{\mathrm{Tr}}\{e^{-\beta (\hat{\mathcal H} - \mu \hat{\mathcal N})} \hat{\mathrm P}^{(m)}_{\bs n}\}}{{\mathcal Z}(\beta,\mu)}, \quad
\label{eq_R_def}
\eeqn
where $\hat{\mathrm P}^{(m)}_{\bs n}$ is the projector onto these states and ${\mathcal Z}(\beta,\mu) = {\mathrm{Tr}}\{e^{-\beta (\hat{\mathcal H} -\mu\hat{\mathcal N})}\}$ is the path integral of the system.
The traces are taken over all the states of the Hilbert space. Inessential volume factors are neglected.

The generating function $K_{\bs n}$, associated with the angular momentum distribution around the axis ${\bs n}$, is defined by the Fourier transform of the relative contribution~\eq{eq_R_def}:
\beqn
e^{-K_{\bs n}(\beta,\mu; \chi)} = \sum_{m \in {\mathbb Z}} e^{i \chi m} \, R_{\bs n}(\beta,\mu; m)\,,
\label{eq_gen_functional}
\eeqn
where the compact quantity $\chi \in (-\pi,\pi]$ is the conjugated variable with respect to the momentum $m$. 

Reversing Eq.~\eq{eq_gen_functional} we get the relative contribution coming from the states with the fixed angular momentum~$m$:
\beqn
R_{\bs n}(\beta,\mu; m) = \int_{-\pi}^\pi \frac{d \chi}{2 \pi} \frac{{\mathcal Z}(\beta,\mu; \chi{\bs n}/\beta) }{{\mathcal Z}(\beta,\mu)}  e^{-i \chi m}, 
\label{eq_R}
\eeqn
where 
\beqn
{\mathcal Z}(\beta,\mu; {\bs \Omega}_I) = {\mathrm{Tr}} \left\{e^{ - \beta(\hat {\mathcal H} - \mu \hat {\mathcal N} - i\hat {\bs \Omega}_I \cdot \hat{\bs {\mathcal J}})}\right\}\,,
\label{eq_Z_Omega_I}
\eeqn
is a partition function in which the states carrying the angular momentum $m$ are weighted by the phase $e^{i \beta {\bs \Omega}_I \cdot \hat{\bs {\mathcal J}}}$. Here $\hat{\bs {\mathcal J}}$ is the total angular momentum operator which, for a particle with a nonzero spin, also includes the spin contribution. For our spin-0 bosonic states $|m\rangle$ with the definite value of $m$, one has $e^{i \beta {\bs \Omega}_I \cdot \hat{\bs {\mathcal J}}} |m\rangle = e^{i\beta \Omega_I m} |m\rangle$ and, therefore, we recover the phase factor in Eq.~\eq{eq_R}. 

The ``rotated'' partition function 
\beqn
{\mathcal Z}(\beta,\mu; {\bs \Omega}_I) = \sum_{m \in {\mathbb Z}} e^{i \beta \Omega_I m} Z_m(\beta,\mu)\,,
\label{eq_Z_sum_Omega_I}
\eeqn
is the path integral in Euclidean time with the rotwisted boundary conditions in the imaginary time~\eq{eq_rotwisted}. Notice that the angle $\beta \Omega_I \equiv \chi$, which determines the phase $e^{i \chi m}$ attributed to different $m$ sectors in the path integral~\eq{eq_Z_sum_Omega_I}, is nothing but the statistical angle~\eq{eq_chi}! Moreover, we immediately arrive to the conclusion that the partition function~\eq{eq_Z_sum_Omega_I} can be represented as a sum over contributions ${\mathcal Z}_m(\beta,\mu)$ coming from different ``topological sectors'' characterized by the total angular momentum with respect to the axis ${\bs n} = {\bs \Omega}_I/\Omega_I$. Evidently, ${\mathcal Z}(\beta,\mu) \equiv {\mathcal Z}(\beta,\mu; \Omega_I = 0)$.

Consequently, the generating function for the relative contributions of the angular-momentum states can be written as the ratio of two partition functions,
\beqn
e^{- K_{\bs n}(\beta,\mu; \Omega_I)} \equiv 
\avr{e^{i \beta \hat {\bs \Omega}_I \cdot \hat{\bs {\mathcal J}}}}= \frac{{\mathcal Z}(\beta,\mu; {\bs \Omega}_I)}{{\mathcal Z}(\beta,\mu)}\,,
\label{eq_K}
\eeqn
corresponding to the rotwisted~\eq{eq_rotwisted} and standard~\eq{eq_fields_boundary} boundary conditions of the same theory. The quantity~\eq{eq_K} has a well-defined thermodynamic limit.

Therefore, at the level of the path integral, the rotwisted boundary conditions~\eq{eq_rotwisted} produce the generating functional~\eq{eq_K} for the states with the definite projection of the angular momentum number $m$. The statistical angle $\chi$, related to the imaginary angular rotation frequency~$\Omega_I$ via Eq.~\eq{eq_chi}, plays a role of the conjugate variable with respect to the angular momentum~$m$.

The partition function for a field theory can also be formulated as a path integral over the fields~\cite{Kapusta:2006pm}. For a scalar field theory in $d+1$ dimensional time, the finite-temperature partition function~\eq{eq_Z_Omega_I} can be written as:
\beqn
& & {\mathcal Z}(\beta,\mu; {\bs \Omega}_I) = \int D \pi \int_{\mathrm{rotwisted}({\bs \Omega}_I)} D \phi 
\exp \biggl\{ \int_0^\beta \!\! d \tau \nonumber  \\
& & 
\ \times \int d^d x \biggl[ i \pi(x) \frac{\partial \phi(x)}{\partial \tau} {-} {\mathcal H}\bigl( \pi, \phi \bigr){+} \mu {\mathcal N}\bigl( \pi, \phi \bigr)\biggr] \biggr\}.  \qquad
\label{eq_Z_rotwisted_path}
\eeqn
The term ``rotwisted$({\bs \Omega}_I)$'' indicates that the integration is performed over the scalar field $\phi(x)$ subjected to the rotwisted boundary condition~\eq{eq_phi_rotation} with the angle ${\bs \chi} = \beta {\bs \Omega}_I$. The integration over the conjugate momentum  $\pi(x)$ is not restricted.

For a fermionic system, the angular momentum in the phase factors of Eqs.~\eq{eq_gen_functional}, \eq{eq_R}, and \eq{eq_Z_sum_Omega_I} should be shifted, $m \to m + 1/2$. This change reflects the anti-periodicity of the fermionic fields under $2\pi$ spatial rotations which also enters the corresponding rotwisted boundary conditions~\eq{eq_psi_rotation}. The partition function can be formulated similarly to Eq.~\eq{eq_Z_rotwisted_path}.

\subsection{Commutation relations for ninions}

The standard imaginary-time formalism implies that the bosonic and fermionic) fields are periodic (anti-periodic) functions of the imaginary time~\eq{eq_fields_boundary}. This requirement is closely related to the spin-statistics theorem, which implies, in 3+1 dimensions, that bosonic fields~$\phi$ (fermionic fields $\psi$) are represented by commuting (anti-commuting) operators, respectively: 
\beqn
[\phi(x) ,\phi (x')] = 0, \quad \{\psi(x), \psi(x')\} = 0\,.
\label{eq_commutation}
\eeqn
Here equal times ($t=t'$) and different spatial coordinates (${\bs x} \neq {\bs x}'$) are implictly imposed. 

Since the spin-statistical properties in (3+1)d determine the boundary conditions~\eq{eq_fields_boundary} in the 4d Euclidean space with imaginary time, one could naturally ask ourselves whether the rotwisted boundary conditions~\eq{eq_rotwisted} bring a modification to the commutation relations between the field operators, and thus determine a new statistics similar to one of the anyons in (2+1)d?

To address this point, we notice that the path integral \eq{eq_Z_Omega_I} implies that the evolution of the imaginarily rotating system is governed by the Hamiltonian~\eq{eq_H_Omega_I}. For simplicity, we consider a zero-density limit, $\mu = 0$. The thermal Green's function is~\cite{das1997finite,Mustafa:2022got},
\begin{eqnarray}
 G_\beta(\tau,\tau') & = & \avr{{\mathcal T} \left[ \Phi_H(\tau) \Phi_H(\tau') \right]}_\beta 
 \label{eq_Green_function_0} \\
& = & {\mathrm{Tr}}\, \left ( \hat\rho \, {\mathcal T}  \left[ \Phi_H(\tau) \Phi_H(\tau') \right]\right),  \nonumber 
\end{eqnarray}
where we introduced, for brevity, the statistical operator:
\beqn
{\hat \rho} \equiv {\hat \rho}_{\beta,{\bs \Omega}_I} = {\cal Z}^{-1}(\beta) e^{-\beta \hat{\mathcal H}_{{\bs \Omega}_I}}\,. 
\label{eq_statistical_operator}
\eeqn
The Green's function~\eq{eq_Green_function_0} also involves the imaginary-time ordering operator,
\begin{eqnarray}
 {\mathcal T} \left[\Phi_H(\tau) \Phi_H(\tau') \right] & = & \phantom{\pm} \Theta(\tau-\tau') \Phi_H(\tau) \Phi_H(\tau') \\
  & & \pm \Theta(\tau'-\tau) \Phi_H(\tau') \Phi_H(\tau)\,, \nonumber
\end{eqnarray}
where the upper and lower signs correspond to bosonic and fermionic field $\Phi$, respectively. 

In Eq.~\eq{eq_Green_function_0}, the imaginary-time evolution of the field, 
\beqn
\Phi_H(\tau) = e^{\tau \hat{\mathcal H}_{{\bs \Omega}_I}} \, \Phi \, e^{-\tau\hat{\mathcal H}_{{\bs \Omega}_I}}\,, 
\eeqn
is determined by the Hamiltonian~\eq{eq_H_Omega_I}. The imaginary time variables $\tau$ and $\tau'$ are restricted to the principal interval: $0 < \tau, \tau < \beta$.

The standard chain of the Kubo-Martin-Schwinger (KMS) relation~\cite{das1997finite,Mustafa:2022got} gives us:
\beqn
& & G_\beta({\bs x}, {\bs x}';\tau,\tau') {\biggl|}_{\tau >\tau'} 
= {\mathrm{Tr}} \left({\hat \rho} \, {\mathcal T} \left[\Phi_H({\bs x},\tau) \Phi_H({\bs x}',\tau') \right]\right) \nonumber \\
& = & {\mathrm{Tr}}\,\left [ {\hat \rho} \, \Phi_H({\bs x},\tau) \Phi_H({\bs x}',\tau') \Theta(\tau-\tau') \right] \nonumber \\
& = & {\mathrm{Tr}} \left [ \Theta(\tau-\tau')  \Phi_H({\bs x}',\tau') \, \hat\rho \, \Phi_H({\bs x},\tau)
 \right] \nonumber \\
& = & {\mathrm{Tr}} \left [ \Theta(\tau-\tau') e^{-\beta \hat{\mathcal H}_{{\bs \Omega}_I}} e^{\beta \hat{\mathcal H}_{{\bs \Omega}_I}} \Phi_H({\bs x}',\tau') \, {\hat \rho} \, \Phi_H({\bs x},\tau)
 \right] \nonumber \\
& = & {\mathrm{Tr}} \left [ \Theta(\tau-\tau')\, \hat\rho\, \Phi_H({\hat R}_{{\bs \chi}} {\bs x}',\tau'+\beta)  \Phi_H({x},\tau)
 \right] \nonumber \\
& = & \pm {\mathrm{Tr}}\,\left [\hat\rho \, \Theta(\tau - \tau') \Phi_H({\bs x},\tau) \Phi_H({\hat R}_{{\bs \chi}} {\bs x}',\tau'+\beta) 
 \right] \nonumber \\
& = & \pm {\mathrm{Tr}}\left({\hat \rho} \, {\mathcal T} \left[\Phi_H({\bs x},\tau) \Phi_H({\hat R}_{{\bs \chi}} {\bs x}',\tau'+\beta) \right]\right) \nonumber \\
& \equiv & \pm \ G_\beta\left({\bs x}, {\hat R}_{{\bs \chi}} {\bs x}';\tau ,\tau'+\beta\right),
\label{eq_chain_Green}
\eeqn
where we inserted the unit operator $1 =  e^{-\beta \hat{\mathcal H}_{{\bs \Omega}_I}} e^{\beta \hat{\mathcal H}_{{\bs \Omega}_I}}$, took into account the commutation relations~\eq{eq_commutation}, and used the cyclic properties of the trace. For definiteness, we took $\tau > \tau'$ (see also Ref.~\cite{Ambrus:2021eod} for the case of real rotation). 

In deriving Eq.~\eq{eq_chain_Green}, we also used the imaginary-time evolution of the state: 
\beqn
\Phi_H({\hat R}_{{\bs \chi}} {\bs x},\tau+\beta)= e^{\beta \hat{\mathcal H}_{{\bs \Omega}_I}} \Phi_H({\bs x},\tau) e^{-\beta \hat{\mathcal H}_{{\bs \Omega}_I}}\,,
\label{eq_evolution_H_Omega_I}
\eeqn
which follows directly from the form of the Hamiltonian~\eq{eq_H_Omega_I}. The evolution~\eq{eq_evolution_H_Omega_I} translates the state for one step along the imaginary time, $\tau \to \tau + \beta$, and simultaneously rotates it by the angle \eq{eq_chi}, ${\bs x} \to {\hat R}_{{\bs \chi}} {\bs x}$ about the spatial axis ${\bs n} = {\bs \Omega}_I/\Omega_I$. For fermions, Eq.~\eq{eq_evolution_H_Omega_I} also includes rotation in the spinor space~\eq{eq_psi_rotation} which is not shown explicitly.

The KMS relations set the boundary conditions for the fields along the compactified imaginary-time direction~\cite{das1997finite,Mustafa:2022got}. In our case, the chain of the relations~\eq{eq_chain_Green} for the rotwisted boundary conditions~\eq{eq_rotwisted} gives us the standard commutation relations~\eq{eq_commutation} for bosonic, $\Phi = \phi$ and fermionic, $\Phi = \psi$, fields. Therefore, we conclude that under imaginary rotation, the bosons remain bosons, and the fermions remain fermions: no spin-statistical transformation for these particles occurs. However, as we already demonstrated for a generic system with imaginary rotation, the occupation numbers at different energy levels do not correspond, at thermal equilibrium, to bosonic or fermionic statistics. 

Thus, the ninion system has an exotic statistical property that is different from the standard fractional statistics: the form of the occupation numbers for bosonic and fermionic particles dependent on the level they occupy. In general, the level-dependent thermal distribution~\eq{eq_n_chi} is related, via Eq.~\eq{eq_chi_bos_ferm}, to a continuous statistical parameter~$\chi$ which interpolates between different types of statistical distributions. In the limiting cases, the thermal distribution takes either the standard bosonic or fermionic form.

\subsection{Non-Hermiticity, $\bs {PT}$-invariance, and unitarity of imaginary rotation}

As we have already noticed, the form of the statistical operator~\eq{eq_statistical_operator} implies that the evolution of the imaginary rotating system is described by the ``rotwisted'' Hamiltonian~\eq{eq_H_Omega_I}. One can easily prove that, due to the presence of the purely imaginary term, this Hamiltonian is a non-Hermitian operator, $\hat{\mathcal H}_{\Omega_I}^\dagger \neq \hat{\mathcal H}_{\Omega_I}$, which could potentially lead to dissipation or instability in the system. However, it is somewhat amusing to recognize that so far we did not find any trace of these unwanted artificial features in the imaginary-time formulation of thermodynamics.

The reason behind this unusual property can be rooted in the invariance of the Hamiltonian~\eq{eq_H_Omega_I} under the common action of the parity transformation, $P: \ {\bs x} \to - {\bs x}$, and the time reversal, $T: \ t \to - t$. Indeed, the total angular momentum $\hat {\bs {\mathcal J}}$ is a parity even ($P \hat {\bs {\mathcal J}} P = \hat {\bs {\mathcal J}}$) and time-reversal odd ($T \hat {\bs {\mathcal J}} T = - \hat {\bs {\mathcal J}}$) operator. Given the anti-linearity of the time reversal, $T\, i \, T = - i$, one finds that the rotwisted term of the Hamiltonian~\eq{eq_H_Omega_I} is a $PT$-invariant operator:
\beqn
PT(i\hat {\bs \Omega}_I \cdot \hat{\bs {\mathcal J}}) PT = i\hat {\bs \Omega}_I \cdot \hat{\bs {\mathcal J}}\,.
\eeqn
In the $PT$-unbroken regime, the $PT$-symmetric non-Hamiltonians share the properties of the usual Hermitian Hamiltonians: they have a real-valued energy spectrum, describe the unitary evolution system and exhibit a well-defined thermodynamic limit~\cite{Bender:1998ke}. Perhaps, we observe a somewhat similar effect in the imaginary time formalism: despite the non-Hermiticity of the Hamiltonian~\eq{eq_H_Omega_I}, the imaginary rotation appears to be a well-defined concept in thermodynamics.

\subsection{Analogies with Quantum Chromodynamics}

\subsubsection{$\theta$-angle in QCD in the path-integral formulation}

The last term of the rotwisted Hamiltonian~\eq{eq_H_Omega_I} shares similarity with the $\theta$ term in the theory of strong fundamental interactions, Quantum Chromodynamics (QCD). The $\theta$ term in QCD has a topological origin, 
\beqn
{\mathcal J}^\theta_{\mathrm{QCD}} = - \theta q(x)\,,
\eeqn
where
\beqn
q(x) = \frac{g^2}{16\pi^2} {\mathrm{Tr}} \left[ G_{\mu\nu} {\tilde G}^{\mu\nu} \right]\,,
\label{eq_top_density}
\eeqn
is the topological charge density expressed via the gluonic field strength tensor $G_{\mu\nu} = \partial_\mu A_\nu - \partial_\nu A_\mu + i [A_\mu,A_\nu]$ and its dual ${\tilde G}_{\mu\nu}=\frac{1}{2}\epsilon_{\mu\nu\rho\sigma} G^{\rho\sigma}$. The integral over the topological density~\eq{eq_top_density},
\beqn
\nu[A] = \int d^4 x \,w(x) \in {\mathbb Z}\,,
\label{eq_QCD_nu}
\eeqn
is an integer which determines the topological winding number $\nu$ of the gluonic configuration $A_\mu$. The potential existence of the $\theta$ term in the theory leads to ``strong $CP$-problem'' (for a review, see Ref.~\cite{RevModPhys.82.557}).

The vacuum structure of QCD depends on the distribution of the topological winding number~\eq{eq_QCD_nu} which determines important properties of QCD vacuum~\cite{callan1976structure,leutwyler1992spectrum}. The Euclidean action of QCD with the $\theta$ term is
\beqn
S^{\mathrm{QCD}}(\theta) & = S^{\mathrm{QCD}}_0 - i \theta \nu[A]\,,
\label{eq_S_QCD_theta}
\eeqn
where 
\beqn
S^{\mathrm{QCD}}_0 = S_G + \int d^4 x \bar{q}\left(-i \Dirac + M \right)q\,,
\label{eq_S_QCD}
\eeqn
is the standard Euclidean action which contains the gauge Yang-Mills term, 
\beqn
S_G = \int d^4 x \frac{1}{2}{\mathrm{Tr}} \left[G_{\mu\nu} G_{\mu\nu} \right] \,,
\eeqn
and the quark sector with the spinor fields $q$, the mass matrix $M$, and the covariant derivative $\Dirac = \gamma^\mu D_\mu(A)$.

The QCD partition function involves the sum over all gauge-field configurations which are characterized by the winding number $\nu$. Therefore, the partition function is a sum over different topological sectors $\nu$ weighted with the phase which depends on the $\theta$ angle:
\beqn
{\mathcal Z}^{\mathrm{QCD}}(\theta) = \sum_{\nu \in {\mathbb Z}} e^{i\theta\nu} {\mathcal Z}^{\mathrm{QCD}}_\nu\,.
\label{eq_Z_QCD_sum} 
\eeqn
In the partition function,
\beqn
{\mathcal Z}^{\mathrm{QCD}}_\nu = \int [{\mathrm D}A_\mu ]_\nu \ {\mathrm{det}} \left(-i \Dirac + M\right) e^{-S_G}\,,
\label{eq_QCD_at_nu}
\eeqn
the integration over the gluons $A_\mu$ restricted to the sector with the fixed topological winding number $\nu$. The functional determinant in Eq.~\eq{eq_QCD_at_nu} comes as a result of the integration over the quark fields $q$.

One notices the striking similarity between the partition function of QCD~\eq{eq_Z_QCD_sum} and the partition function of the generic field theory~\eq{eq_Z_sum_Omega_I} subjected to the rotwisted boundary conditions~\eq{eq_rotwisted}: both functions are determined by the sums of the respective topological sectors. The similarity is enforced by the observation that the topological term in Eq.~\eq{eq_S_QCD_theta} is an analog of the last term in the rotwisted Hamiltonian~\eq{eq_H_Omega_I} which determines the imaginary rotation. The $\theta$-angle parameter in QCD is analogous to the statistical parameter $\chi$ which defines the rotwisted boundary conditions. Moreover, the integer-valued topological winding number $\nu$ corresponds to the angular momentum quantum number~$m$ (in a certain sense, the analogy is even more expressive in the physical sense since the angular momentum counts how fast the system ``winds'' about the given axis).

\subsubsection{$\theta$-vacuum in QCD}

The analogy between topology in QCD and imaginary rotation goes further. Below we consider the QCD $\theta$-vacuum encoded in the topological properties of the gluon fields. We consider Yang-Mills theory which describes the gluonic sector of the theory. 

The lowest-energy state of the classical Yang-Mills theory is an infinitely degenerate state due to the presence of the infinitely many classical minima~\cite{rubakov2009classical}. The minima are topologically inequivalent gluon configurations labeled by the topological number $n \in {\mathrm Z}$. This number is usually associated with the difference between so-called Chern-Simons numbers $n_{\mathrm{CS}} \equiv N_{\mathrm{CS}}[A]$ of the given vacuum state and a ``fiducial'' vacuum state. 

In a semi-classical Yang-Mills theory, the transition between vacuum states is described by instantons~\cite{belavin1975pseudoparticle} which are gluonic configurations $A_\mu$ that are interpreted as tunneling events. An instanton event changes the topological number of the state, $n_{\mathrm{CS}} \to n_{\mathrm{CS}} + \nu$, by the integer-valued topological number~\eq{eq_QCD_nu} $\nu = \nu[A]$ of the instanton. 

In quantum theory, the tunneling between neighboring vacuum states removes the classical degeneracy of the ground state. The quantum ground state is now described by the $\theta$ vacuum:
\beqn
\left| \theta \right\rangle = \sum_{\nu \in {\mathbb Z}} e^{i \theta \nu} \left| \nu \right\rangle\,.
\label{eq_theta_vacuum}
\eeqn
The value of $\theta$ labels different vacuum states and, thus, removes the classical degeneracy in the quantized non-Abelian Yang-Mills gauge theory. The $\theta$ angle serves as an independent physical parameter of the theory~\cite{rubakov2009classical}.

This situation, of course, is not pertinent only to Yang-Mills theory. Examples include quantum mechanics of a particle in a periodic potential, where the role of $\theta$ is played by a quasimomentum in the Bloch phase, and a charged particle on a ring, where the particle states feel the Aharonov-Bohm phase~\cite{rubakov2009classical}. In analogy with quantum mechanical properties of a harmonic oscillator, the $\theta$-vacuum state~\eq{eq_theta_vacuum} can be represented, up to a global phase, as a sum of the extended coherent states~\cite{klauder1993extended}: 
\beqn
\left| \nu \right\rangle_\theta = e^{i \theta {\hat N}_{\mathrm{CS}}} \left| \nu \right\rangle, 
\label{eq_coherent_QCD}
\eeqn
with distinct vacua $\left| \nu \right\rangle$ serving as fiducial vectors. Below, we will show that the ninions are the coherent states of the rotation operators similar to the topological coherent states in QCD~\eq{eq_coherent_QCD}.

\subsection{Coherent states and imaginary rotation}

Coming back to the rotwisted boundary conditions, we define the coherent rotational ``$\chi$-state'' in similarity with the $\theta$-vacuum in QCD~\eq{eq_theta_vacuum}: 
\beqn
\left| \alpha ; \chi \right\rangle_{\bs n} {=}\! \sum_{m \in {\mathcal M}_\alpha} e^{i \chi m} \left| \alpha, m \right\rangle_{\bs n} {\equiv}\! \sum_{m \in {\mathcal M}_\alpha} e^{i \chi {\hat{\mathcal J}}_{\bs n}} \left| \alpha, m \right\rangle_{\bs n}\!,\quad
\label{eq_vacuum_theta_I}
\eeqn
where $\left|\alpha, m \right\rangle_{\bs n}$ is a vector in the Hilbert space representing a state which carries the angular momentum $m$ with respect to a fixed axis $\bs n$ and ${\hat{\mathcal J}}_{\bs n} \equiv {\bs n} \cdot {\hat{\bs {\mathcal J}}}$ is the operator of projection of the angular momentum onto the axis~$\bs n$.  As usual, the collective index $\alpha$ denotes the quantum numbers other than~$m$. The set ${\mathcal M}_\alpha$ in the sum~\eq{eq_vacuum_theta_I} incorporates all values of the angular momenta $m$ with the same value of energy, $\varepsilon_{\alpha,m_1} = \varepsilon_{\alpha,m_2} = \varepsilon_\alpha$ for $m_{1,2} \in {\mathcal M}_\alpha$. Therefore, the state \eq{eq_vacuum_theta_I} is a coherent state given by the superpositions of all energy-degenerate angular momentum modes:
\beqn
\hat{\mathcal H} \left| \alpha ; \chi \right\rangle_{\bs n} =  \varepsilon_\alpha \left| \alpha ; \chi \right\rangle_{\bs n}\,.
\eeqn

In the absence of degeneracy, the state~\eq{eq_vacuum_theta_I} is represented by a coherent state of the rotation operator:
\beqn
\left| \alpha ; \chi \right\rangle_{\bs n} = e^{i \chi {\hat{{\mathcal J}}}_{\bs n}} \left| \alpha, m \right\rangle_{\bs n},
\label{eq_coherent_chi}
\eeqn
where the state $\left| \alpha, m \right\rangle_{\bs n}$ serves as a fiducial vector in the Hilbert space. 

In quantum optics, the vectors~\eq{eq_coherent_chi} are known as the ``coherent spin states''~\cite{radcliffe1971some,arecchi1972atomic}: the quantum number $m$ is associated with the spin of a particle $s_z$ and the operator ${\hat{{\mathcal J}}}_{\bs n}$ becomes the operator of the spin projection ${\hat S}_z$ on the $z$ axis, with ${\bs n} \equiv {\bf e}_z$. In the context of imaginary rotation, the state~\eq{eq_coherent_chi} is an elementary ninion state in which the coherence parameter ~$\chi$ coincides exactly with the statistical parameter $\chi$ encoded in the rotwisted boundary conditions along the imaginary time direction~\eq{eq_rotwisted}.

Consider now the case of free quantum theory which we studied earlier. The eigenenergies of a free boson~\eq{eq_energy_scalar} are infinitely degenerate with respect to the angular momentum $m$ similarly to the infinitely degenerate ground states of classical Yang-Mills theory that are labeled by the topological index $\nu$. Thus, the rotational $\chi$-state~\eq{eq_vacuum_theta_I} for the free scalar theory can be written in the following suggestive form:
\beqn
\left| p_\rho, p_z ; \chi \right\rangle_{\bs n} = \sum_{m \in {\mathbb Z}} e^{i \chi m} \left| m ; p_\rho, p_z  \right\rangle_{\bs n}\,,
\label{eq_p_Omega}
\eeqn
where the vector $\left| m ; p_\rho, p_z  \right\rangle_{\bs n}$ represents the state~\eq{eq_wave_function_0} and the vector $\bs n$ points along the $z$ axis. For the ground state with a vanishing momentum, $p_\rho = p_z = 0$, the ground state can be called the $\chi$-vacuum state~\eq{eq_vacuum_theta_I} in analogy with the $\theta$-vacuum in QCD~\eq{eq_theta_vacuum}.

Extending further the analogy with the physics of QCD, let us notice that the instantonic transition from one vacuum $\ket{\nu}$ to the other, $\ket{\nu+1}$, can be represented as a change of the phase of the $\theta$-vacuum by $e^{i \theta}$. Therefore, this phase factor is the eigenfunction of the corresponding instanton transition operator~\cite{rubakov2009classical}. 

To maintain the above analogy for the imaginary rotation, let us introduce the operator $\hat{\mathsf T}$ which increases the angular momentum of the rotating system by one, $\hat{\mathsf T} {\ket{m}}_{\bs n} = {\ket{m+1}}_{\bs n}$. In the free theory, the infinite degeneracy of the energy states~\eq{eq_energy_scalar} implies that the Hamiltonian and the shift operator commute with each other, $[\hat{\mathcal H}, \hat{\mathsf T}] = 0$. Therefore, these operators share the eigenstates. The statistical parameter $\chi$ defines the eigenvalue of the angular-momentum-increasing operator, 
\beqn
\hat{\mathsf T} \left| p_\rho, p_z ; \chi \right\rangle_{\bs n} = e^{i \chi} \left| p_\rho, p_z ; \chi \right\rangle_{\bs n}\,.
\eeqn
It is the pure phase, $e^{i\chi}$, similarly to the eigenvalue $e^{i\theta}$ of the instanton operator acting on the QCD $\theta$-vacuum~\cite{rubakov2009classical}.

If the field is interacting with a background potential, then the infinite degeneracy of the states with respect to the angular momentum $m$ is partially lifted out (an example is given in Section~\ref{sec_examples_transmutation_bosons} on page~\pageref{sec_examples_transmutation_bosons}). Therefore, for interacting systems, the sum in Eq.~\eq{eq_p_Omega} goes only over the $m$-degenerate eigenenergies, $m \in {\mathcal M}_\alpha$, leading to the series of the degenerate coherent $\chi$-states~\eq{eq_vacuum_theta_I}. 

In summary, the ninions can be associated with the coherent angular momentum states~\eq{eq_coherent_chi}. In a free theory, such states possess an infinite degeneracy which makes their ground state similar to the $\theta$-vacuum of QCD.

\section{Conclusions}

We discussed quantum field theories in thermal equilibrium in the modified Euclidean imaginary time formalism. The standard (anti-)periodicity of fields in the compactified imaginary time direction is generalized to the ``rotwisted'' boundary condition~\eq{eq_rotwisted} which supplements the advance along the imaginary time with a global spatial rotation (illustrated in Fig.~\ref{fig_cylinder}). We demonstrated that the fields satisfying such a condition --which formally corresponds to rotation with an imaginary angular momentum-- evolve according to a $PT$-symmetric non-Hermitian Hamiltonian, which points to the unitary evolution of the underlying physical system. 

The rotwisted boundary conditions describe new excitations that we call ``ninions'' (the etymology of this word is given in footnote~\ref{footnote} on page \pageref{footnote}). The ninions have intriguing statistical features as the functional dependence of the ninion occupation number on temperature, and chemical potential depends on the quantum eigenstate that this particle occupies. As the functional form of the occupation number determines the statistical distribution of particles in the ensemble, the ninion possesses variable, level-dependent statistics, different from bosons and fermions. The ninions exist in three spatial dimensions and are also different from anyons. 

We gave an explicit example of a ninionic system for which, at a certain eigenstate, the particle occupation number is given by the Bose-Einstein statistics, while at another eigenstate, the same particle becomes a fermionic ghost. Such a ghost is described by the Fermi-Dirac statistics but enters the system's free energy with the wrong, ``bosonic'' sign. 

The ninionic occupation number~\eq{eq_n_chi} is labeled by the continuous number related to the statistical angle $\chi$. This angle enters the rotwisted boundary condition~\eq{eq_chi_bos_ferm} and can be incorporated in the path integral formalism~\eq{eq_Z_rotwisted_path}. The ground state of free ninions shares similarity with the $\theta$-vacuum in Quantum Chromodynamics, with the topological $\theta$ angle playing the role of the statistical parameter $\chi$ while the topological QCD charge is associated with the angular momentum. We argue that the rotwisting (statistical) angle $\chi$ is an independent coupling in thermodynamics in the same sense as the $\theta$ angle is an independent coupling in QCD.

Interestingly, certain ninionic states appear to be ghosts that possess negative pressure~\eq{eq_ghost_thermodynamics}. This property could make a ninionic state an exciting candidate for cosmological dark energy, which could play a role in the observed acceleration of the Universe's expansion.

We also demonstrated that the dependence of thermodynamic characteristics of noninteracting ninions on the statistical angle $\chi$ in thermodynamic limit is determined by the Thomae function~\eq{eq_Thomae} which is discontinuous at every point. This non-analyticity implies a no-go theorem on the absence of analytical continuation between real and imaginary rotation if the latter is introduced via the rotwisted boundary.

A concrete experimental realization of ninions might exist in systems hosting coherent angular momentum states~\cite{atkins1971angular} and coherent spin states~\cite{radcliffe1971some} that play a significant role in quantum optics~\cite{arecchi1972atomic}. The exotic occupation numbers of ninions may point to the possibility of non-equilibrium but steady states. Furthermore, properties of ninions are also accessible for numerical simulations on Euclidean lattices using standard Monte Carlo methods.  

\acknowledgments
The author is grateful to Victor Ambru\cb{s}, Vladimir Goy, and Stam Nicolis for useful and helpful comments.

\end{document}